\pdfoutput=1
\documentclass{article}
\usepackage{amssymb}
\usepackage{amsmath}
\usepackage{amsfonts}
\usepackage[top=1in, left=1in, bottom=1in, right=1in]{geometry}
\usepackage{graphicx}
\usepackage[dvipsnames,usenames]{color}
\begin{document}

\title{ }

\begin{center}
\textbf{{\Large A route to chaos in the Boros--Moll map }}

\vspace{8mm}

\textsc{{\large Laura Gardini}}

\emph{Department of Economics, Society, Politics (DESP), University of Urbino,
Italy \newline E-mail: laura.gardini@uniurb.it\medskip}

\textsc{{\large V\'{\i}ctor Ma\~{n}osa}}

\emph{Department of Mathematics, Universitat Polit\`{e}cnica de Catalunya,
Spain\newline E-mail: victor.manosa@upc.edu\medskip}

\textsc{{\large Iryna Sushko}}

\emph{Institute of Mathematics, National Academy of Sciences of Ukraine,
Ukraine\newline E-mail: sushko@imath.kiev.ua}
\end{center}

\vspace{7mm}

\begin{center}
{\Large Abstract}\emph{\medskip}
\end{center}

{The Boros-Moll map appears as a subsystem of a Landen transformation
associated to certain rational integrals and its dynamics is related to the
convergence of them. In the paper, we study the dynamics of a one-parameter
family of maps which unfolds the Boros-Moll one, showing that the existence of
an unbounded invariant chaotic region in the Boros-Moll map is a peculiar
feature within the family. We relate this singularity with a specific property
of the critical lines that occurs only for this special case. In particular,
we explain how the unbounded chaotic region in the Boros-Moll map appears.
Special attention is devoted to explain the main contact/homoclinic
bifurcations that occur in the family. We also report some other bifurcation
phenomena that appear in the considered unfolding.}

\section{Introduction}

In this paper we study some dynamical properties of the one-parameter family
of planar maps $G_{h}:\mathbb{R}^{2}\rightarrow\mathbb{R}^{2}$ given by%
\begin{equation}
G_{h}(x,y):=\left(  \frac{h(x+y)+xy+9}{(x+y+2)^{4/3}},\frac{x+y+6}%
{(x+y+2)^{2/3}}\right)  . \label{E-gh}%
\end{equation}
For $h=5$, the map $G_{5}$ appears as a subsystem on an uncoupled Landen
transformation defined in $\mathbb{R}^{5}$ introduced by Boros and Moll in
\cite{BM0}, see also \cite{BM}. Roughly speaking, given a definite integral
depending on several parameters, a Landen transformation is a map on these
parameters that leaves invariant the integral, see \cite[p. 412]{Moll} for a
more precise definition, and \cite{AG} for a historical account of Landen
transformations. Indeed, in the above references it is shown that the
dynamical system defined by
\begin{equation}
\left\{
\begin{array}
[c]{l}%
a_{n+1}=\dfrac{5a_{n}+5b_{n}+a_{n}b_{n}+9}{(a_{n}+b_{n}+2)^{4/3}},\quad
b_{n+1}=\dfrac{a_{n}+b_{n}+6}{(a_{n}+b_{n}+2)^{2/3}},\\
\ \\
c_{n+1}=\dfrac{d_{n}+e_{n}+c_{n}}{(a_{n}+b_{n}+2)^{2/3}},\quad d_{n+1}%
=\dfrac{(b_{n}+3)c_{n}+(a_{n}+3)e_{n}+2d_{n}}{a_{n}+b_{n}+2},\\
\ \\
e_{n+1}=\dfrac{c_{n}+e_{n}}{(a_{n}+b_{n}+2)^{1/3}},
\end{array}
\right.  \label{E-LandenR5}%
\end{equation}
is a Landen transformation of the integral
\[
I(a,b,c,d,e)=\int_{0}^{\infty}{\frac{cx^{4}+dx^{2}+e}{x^{6}+ax^{4}+bx^{2}%
+1}\mathrm{d}x},
\]
which means that $I(a_{n+1},b_{n+1},c_{n+1},d_{n+1},e_{n+1})=I(a_{n}%
,b_{n},c_{n},d_{n},e_{n})$. The map $G_{5}$ is, therefore, the one associated
with the subsystem of the denominator's parameters of the integral. This map
has been investigated from a dynamical view point by Chamberland and Moll in
\cite{ChM}, proving that the set of values of the parameters for which the
integral associated to the 5-dimensional Landen transformation converges,
coincides with the basin of attraction of one of the fixed points of the map
$G_{5}$.

The study of the global dynamics of map $G_{5}$ is a challenging task. The
known facts can be summarized as follows: There is a connected open set which
is the basin of attraction of the fixed point $(x,y)=(3,3)$. Coexisting with
this basin, it appears an invariant set with chaotic dynamics.\footnote{It
cannot be called a chaotic attractor, since the closure of a generic
trajectory looks like the full invariant set, and attracts no other point.}
The boundary between these sets is given by one of the two connected
components of an algebraic curve (given in Eq. \eqref{resolvant}, below) that
is part, together with its preimages, of the stable set of a saddle point that
is also contained in the curve.

In this work we consider the unfolding of $G_{5}$ given by the one-parameter
family \eqref{E-gh} and we describe the route that leads to the appearance of
the chaotic set for $h=5$. In particular we analyze the maps $G_{h}$ for
$h\geq5$, say $h\in\lbrack5,6]$, pursuing the appearance of areas with chaotic
dynamics and their bifurcations, and showing that there is a repelling chaotic
set for $h\gtrsim5$. We relate the existence of the chaotic set for $h=5$ with
the merging of a part of the critical line with the mentioned algebraic
curve.\medskip

{The paper is structured as follows. Section \ref{S2} is devoted to study the
two main final bifurcations that appear in the family: the one that leads to
the appearance of an unbounded chaotic repellor that exists for all the values
of the parameter $h\in(5,\widetilde{h}]$, where $\widetilde{h}\simeq5.032$;
and the final creation of the invariant unbounded chaotic set of $G_{5}$.
These bifurcations are studied in Sections \ref{SS-first-hom-bif} and
\ref{SSfinal}, respectively. }

The previous subsections are devoted to explain some general issues of the
family of maps $G_{h}$. In Section \ref{Ss-PhaseG5} we introduce the notation
concerning the so called \emph{critical lines}, which is one of the main tools
in our analysis, and we recall some of the main features of the phase portrait
of the Boros-Moll map. In Section \ref{ssNS} we prove the existence of three
fixed points for all the values of the parameter $h$, and we determine the
value $h_{NS}$ at which a Neimark-Sacker bifurcation of one of the fixed
points occurs, that starts the sequence of bifurcations studied in the paper
(see Fig.{\ref{Ffg}} and Fig.\ref{1DA}). In Section \ref{ssRegular} we show
that the dynamics of the family of maps for values of $h$ greater than
$h_{NS}$ is regular.

Section \ref{Sroute} is devoted to describe some sequences of bifurcations,
the main ones that appear in the selected route to chaos in the Boros-Moll
map, with special emphasis in the creation and contact bifurcations of chaotic
areas, and homoclinic bifurcations related to snap-back repellors.

An Appendix is included with some considerations regarding the partition of
the phase space in zones related to the number of rank-1 preimages of the
family of maps.

\section{General properties of map $G_{h}$ and main final bifurcations}\label{S2}

\subsection{Structure of the phase space of $G_{5}$}\label{Ss-PhaseG5}

\begin{figure}
[h]
\begin{center}
\includegraphics[scale=0.37]{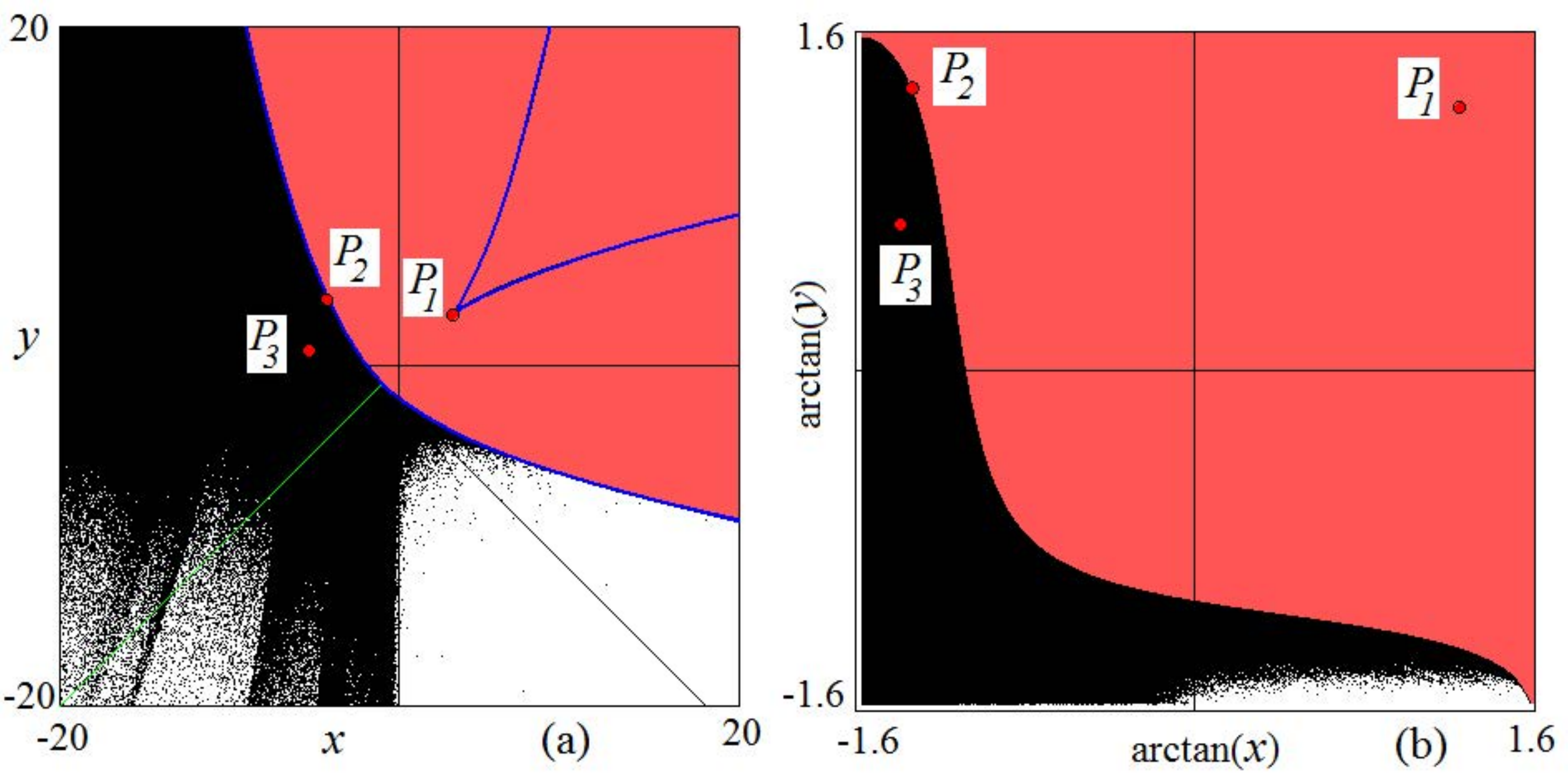}
\caption{Chaotic set (in black) and basin of attraction of $P_{1}=(3,3)$ (in
red) for the Boros-Moll map $G_{5}$. In (a) in the phase plane $(x,y)$. In
(b)\ in the plane scaled as $(\arctan(x),\arctan(y)).$}
\label{h5}
\end{center}
\end{figure}

As mentioned in the Introduction, our aim is to describe how the chaotic set
existing for the map $G_{5}$ is created. In Fig.\ref{h5} we recall the
structure in the phase space.

{The determinant of the Jacobian matrix of any map $G_{h}$ is
\begin{equation}
\det(J(G_{h}))=-\frac{1}{3}\,{\frac{\left(  x+y-6\right)  \left(  x-y\right)
}{\left(  x+y+2\right)  ^{3}}} \label{E-jacobi}%
\end{equation}
and therefore, for any fixed $h\in\mathbb{R}$ the critical curve $LC_{-1}$ of
a map $G_{h}$, is given by the two straight lines
\begin{equation}%
\begin{array}
[c]{l}%
LC_{-1}^{j}:\,y=x,\mbox{ and }\\
LC_{-1}^{jj}:\,x+y-6=0.
\end{array}
\label{E-LC-1}%
\end{equation}
}

{Since $LC_{-1}^{j}$ intersects the line }${(d)}${
\begin{equation}
(d):\,x+y+2=0,\label{(d)}%
\end{equation}
where the denominator of the components of the map vanishes, at the point
}$(x,y)=(-1,-1)${,} the image\footnote{See \cite{BGM} for details related to
some properties of maps with a vanishing denominator.} of $LC_{-1}^{j}$
consists of two unbounded arcs {$LC=G_{h}(LC_{-1}^{j})=LC^{j}\cup LC_{r}^{j}$
which, in the case $h=5$, are the two connected components of the resolvent
curve given by }
\begin{equation}
R(x,y)=-{x}^{2}{y}^{2}+4\,{x}^{3}+4\,{y}^{3}-18\,xy+27=0.\label{resolvant}%
\end{equation}
For $h=5$, the image of the portion of straight line $LC_{-1}^{j}$ from
infinity ${(-}\infty)$ to $(x,y)=(-1,-1)$ gives $LC^{j},$ the arc bounding the
chaotic area in Fig.\ref{h5}, and belongs to the stable set of the saddle
fixed point $P_{2}\simeq(-4.20557,3.95774)$, \cite[Proposition 8]{GLLM}. {The
second branch of $G_{5}(LC_{-1}^{j})$, which is $LC_{r}^{j},$ is the image of
the half line of $LC_{-1}^{j}$ taken from $(-1,-1)$ to infinity }${(+}\infty
)${, and it is also shown in Fig.\ref{h5}a. It includes the fixed point
$P_{1}=(3,3)$ which is an attracting node whose basin is also bounded by
$LC^{j}$.} The third fixed point $P_{3}\simeq(-5.30914,0.83118)$ is an
unstable focus belonging to the chaotic area. {As noticed in \cite{ChM}, the
fact that
\begin{equation}
R(G_{5}(x,y))=\frac{(x-y)^{2}}{(x+y+2)^{4}}R(x,y),\label{E-Rinv}%
\end{equation}
implies that the curve $LC$ is invariant, i.e. $G_{5}^{n}(LC)=LC$ for any
$n>0$. By using the fact that $LC^{j}$ is part of the stable set of the saddle
point $P_{2}$ we also have that the arc $LC^{j}$ is invariant. From this fact,
and from the characterization of the basin of attraction of $P_{1}$ given in
\cite{ChM}, the (seemingly chaotic) area bounded by the curve $LC^{j}$ is also
invariant. In fact, in \cite{Moll02} and \cite[p. 442--444]{Moll} it seems
suggested that such invariant area is exactly a true chaotic area (with dense
periodic points, which are all homoclinic), but this is still to be proved
rigorously.}

\medskip

In order to understand the appearance of the chaotic area we consider the
2-parametric family of maps $G_{g,h}:\mathbb{R}^{2}\rightarrow\mathbb{R}^{2}$
given by
\[
G_{g,h}(x,y):=\left(  \frac{gx+hy+xy+9}{(x+y+2)^{4/3}},\frac{x+y+6}%
{(x+y+2)^{2/3}}\right)  ,
\]
{for which a numerical exploration evidences that unfolding the Boros-Moll map
there are maps with different qualitative behavior. Indeed, one can check that
in the parameter plane ${(g,h)}$ depicted in Fig.~\ref{Ffg} there appear
several regions indicating qualitative differences. The different colors in
this figure represent regions in the parameter space related to the existence
of attracting cycles of different periods.} The yellow region in
Fig.~\ref{Ffg} represents the values of the parameters at which the dynamics
is regular, namely there exist two attracting fixed points, $P_{1}$ in the
positive side and $P_{3}$\ in the negative side, whose basins of attraction
are separated by the stable set of a saddle {fixed point }$P_{2}$.
Differently, {the brown region represents the values of the parameters at
which the only attracting set is }the positive {fixed point} $P_{1}.$ The
existence of these fixed points, as a function of the parameters, is commented
below. The white points denote either existence of a cycle of period higher
than 45 or the existence of a chaotic attractor. Since the positive {fixed
point} $P_{1}$ is always attracting, in all these regions {the attractor
}$P_{1}${ coexists with some other attracting set.
\begin{figure}
[h]
\begin{center}
\includegraphics[scale=0.37]{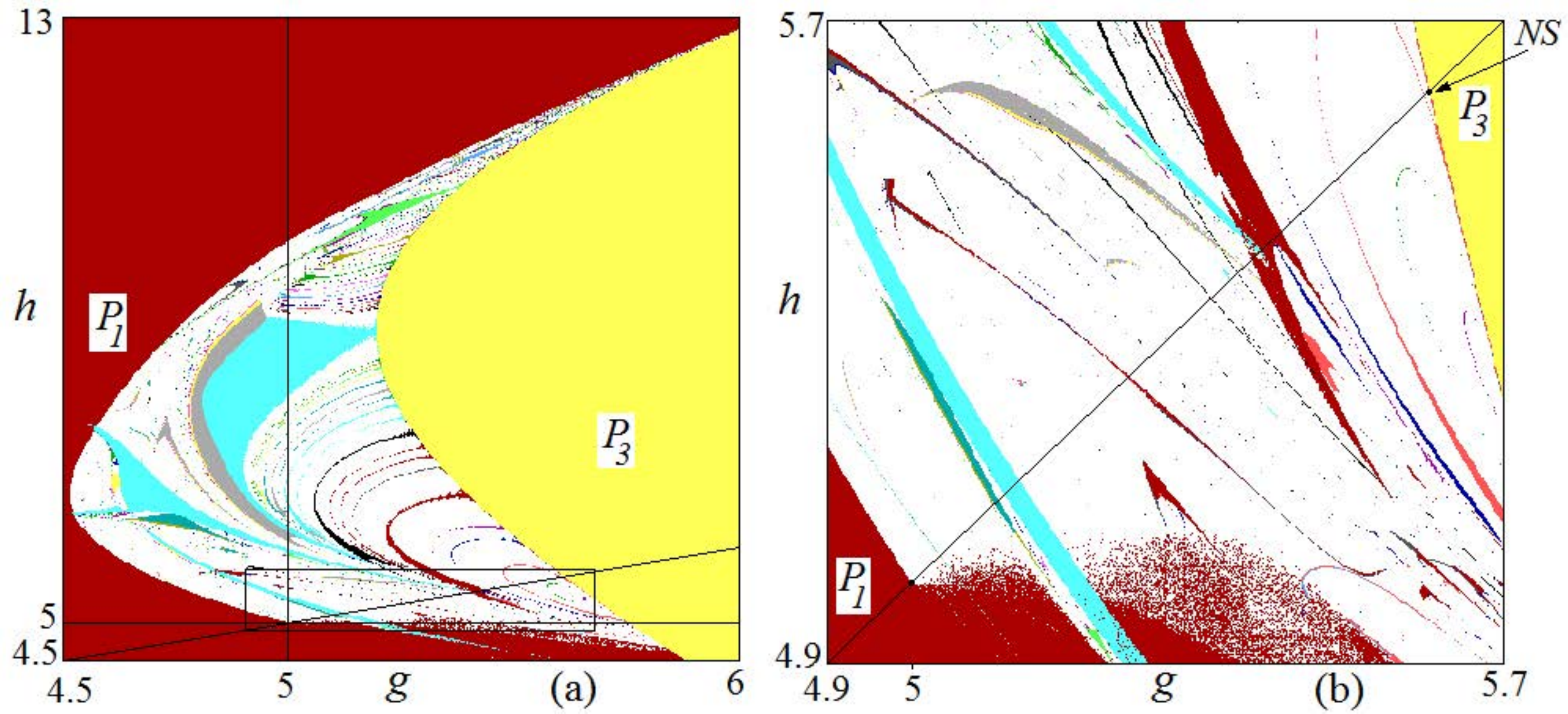}
\caption{In (a) two-dimensional bifurcation diagram related to an i.c. close
to the fixed point $P_{3}.$ In (b) enlargement of the rectangle marked in (a).
}
\label{Ffg}
\end{center}
\end{figure}
To simplify our analysis, which is oriented to understand the dynamics of the
Boros-Moll map, we consider $g=h$, which gives the one-parametric family
\eqref{E-gh}, and we will focus mainly in the interval $h\in\lbrack5,6]$ of
the parameter values. Thus, in the two-dimensional bifurcation diagram the
straight line of equation }${g=h}${ represents the path that we shall follow
in order to describe some of the bifurcations occurring from a regular regime
existing for }${h=6}${ to the chaotic one existing for }${h=5}${. }In fact,
changing the parameter $h$ along the path in Fig.{\ref{Ffg}(b), decreasing
}${h}${, the fixed point }$P_{3}$ becomes an attracting focus and then it
becomes a repelling focus via a Neimark-Sacker (NS from now on, for short)
bifurcation, as commented below, {giving rise to a sequence of bifurcations
that we will summarize in its main features.} {Decreasing }${h}${ we can
comment the appearance/disappearance of attracting }${k}${-cycles for }${k>1}%
${ of different periods or attracting closed invariant curves as well as the
existence of chaotic attractors/repellors.}

\subsection{Fixed points of $G_{h}$ and first Neimark-Sacker bifurcation}\label{ssNS}

We notice that for any $h\in{\mathbb{R}}$, map $G_{h}$ has exactly three
different fixed points. In order to characterize the fixed points of map
$G_{h}$, and following \cite{ChM}, we introduce the auxiliary variable
$m^{3}=x+y+2$, so that the fixed points are characterized by%
\begin{equation}
\left\{
\begin{tabular}
[c]{rl}%
$c_{1}(x,y,m)$ & $:=m^{3}-x-y-2=0,$\\
$c_{2}(x,y,m)$ & $:=-xm^{4}+xy+hx+hy+9=0,$\\
$c_{3}(x,y,m)$ & $:=-ym^{2}+x+y+6=0.$%
\end{tabular}
\ \ \right.
\end{equation}
Isolating $x$ and $y$ from the first and third equations we get:
\begin{equation}
x={\frac{{m}^{5}-{m}^{3}-2\,{m}^{2}-4}{{m}^{2}}}\,\mbox{ and }y={\frac{{m}%
^{3}+4}{{m}^{2}}}.\label{E-fixed-p2}%
\end{equation}
Substituting these expressions in the second equation we obtain that the
variable $m$ must satisfy
\[
c_{4}(m)={m}^{11}-{m}^{9}-3{m}^{8}-h{m}^{7}-3{m}^{6}-2{m}^{5}+\left(
2h-9\right)  {m}^{4}+8{m}^{3}+8{m}^{2}+16=0.
\]
Observe that the term with highest degree in $m$ of $c_{4}(m;h)$ does not
vanish for any $h\in{\mathbb{R}}$, hence no root enters from infinity. On the
other hand, the discriminant of $c_{4}$ is given by
\begin{align*}
&  \Delta_{m}(c_{4}(m;h))=927712935936\,{h}^{14}-29686813949952\,{h}%
^{13}+211982405861376\,{h}^{12}\\
&  +2739735613145088\,{h}^{11}+118044964423729152\,{h}^{10}%
-1417849672260648960\,{h}^{9}\\
&  -3159536320125075456\,{h}^{8}-89793115515877588992\,{h}^{7}%
+2232890826264743510016\,{h}^{6}\\
&  -8668279665625562873856\,{h}^{5}+39121043805342246371328\,{h}^{4}\\
&  -958082327734127689728000\,{h}^{3}+8262772575788649550970880\,{h}^{2}\\
&  -28385116733510952733900800\,h+35844037650190215269056512.
\end{align*}
By using the Sturm method we obtain that $\Delta_{m}(c_{4}(m))>0$ for all
$h\in{\mathbb{R}}$, hence $c_{4}(m)$ has no multiple roots, so that no roots
of $c_{4}(m)$ appear from $\mathbb{C}$ and there is no value of $h$ such that
the roots collide. Finally, using again the Sturm method we have that for
$h=5$, $c_{4}(m)$ has only three distinct real roots. Hence for all
$h\in{\mathbb{R}}$ the polynomial $c_{4}(m)$ has three different real roots,
which give rise to three different fixed points via Eq. \eqref{E-fixed-p2}.

{As already noticed in \cite{ChM} and recalled above, $G_{5}$ has the fixed
points $P_{1}$, $P_{2}$ and $P_{3}$. Since the location of the fixed points
vary continuously with the parameter $h$, we will denote the fixed points of
the maps $G_{h}$ as $P_{1}(h),P_{2}(h)$, and $P_{3}(h)$, not indicating the
dependence on $h$ when it is not necessary.}

\medskip

Since the fixed points $P_{1}$ and $P_{2}$ persist {for $h\in\lbrack5,6]$} as
an attracting node and a saddle, respectively, we are interested in the fixed
point $P_{3}$ which may be attracting or repelling, as we have seen in
Fig.~\ref{Ffg}. So we restrict the scope of our analysis to the proof of a NS
bifurcation that occurs at $h\simeq5.6105116077$. To find the parameter value
where the NS bifurcation takes place we consider the equation $\det
(DG_{h}(x,y))=1$, obtaining
\[
c_{5}(x,y,m):=3\,\left(  x+y+2\right)  ^{3}+\left(  x+y-6\right)  \left(
x-y\right)  =0.
\]
Using the polynomials $c_{1},c_{2}$ and $c_{3}$ that characterize the fixed
points, and taking successive resultants we get that if a NS bifurcation takes
place then the parameter $h$ must be a root of the polynomial
\begin{align*}
&  P_{NS}(h)=582371795533824{h}^{11}-7007247733358592{h}^{10}%
+41393275596177408{h}^{9}\\
&  -557755817005154304{h}^{8}+10296285821493313536{h}^{7}%
-129039497006709473280{h}^{6}\\
&  +970090676944581427200{h}^{5}-4072496100968654438400{h}^{4}%
+8531720267711624773632{h}^{3}\\
&  -7149146865159609778176{h}^{2}%
+6956611779457796014080h-20284715289100324700160
\end{align*}
By using the Sturm method we get that this polynomial has a unique real root
in
\[
\left[  {\frac{808559935549815353}{144115188075855872}},{\frac
{404279967774907677}{72057594037927936}}\right]  .
\]
Hence $h_{NS}\simeq5.6105116077302352$. Now we check that the point $P_{3}(h)$
is a hyperbolic stable focus for $h\gtrsim h_{NS}$, a hyperbolic unstable one
for $h\lesssim h_{NS}$ and non-hyperbolic for $h=h_{NS}$. {In the next
subsection we show that for $h>h_{NS}$ there is coexistence of the two
attracting fixed points $P_{1}$ and $P_{3}$ and their basins of attraction are
separated by the stable set of the saddle $P_{2}$.}

\subsection{Regular regime of the map $G_{h}$}\label{ssRegular}

\begin{figure}
[h]
\begin{center}
\includegraphics[scale=0.37]{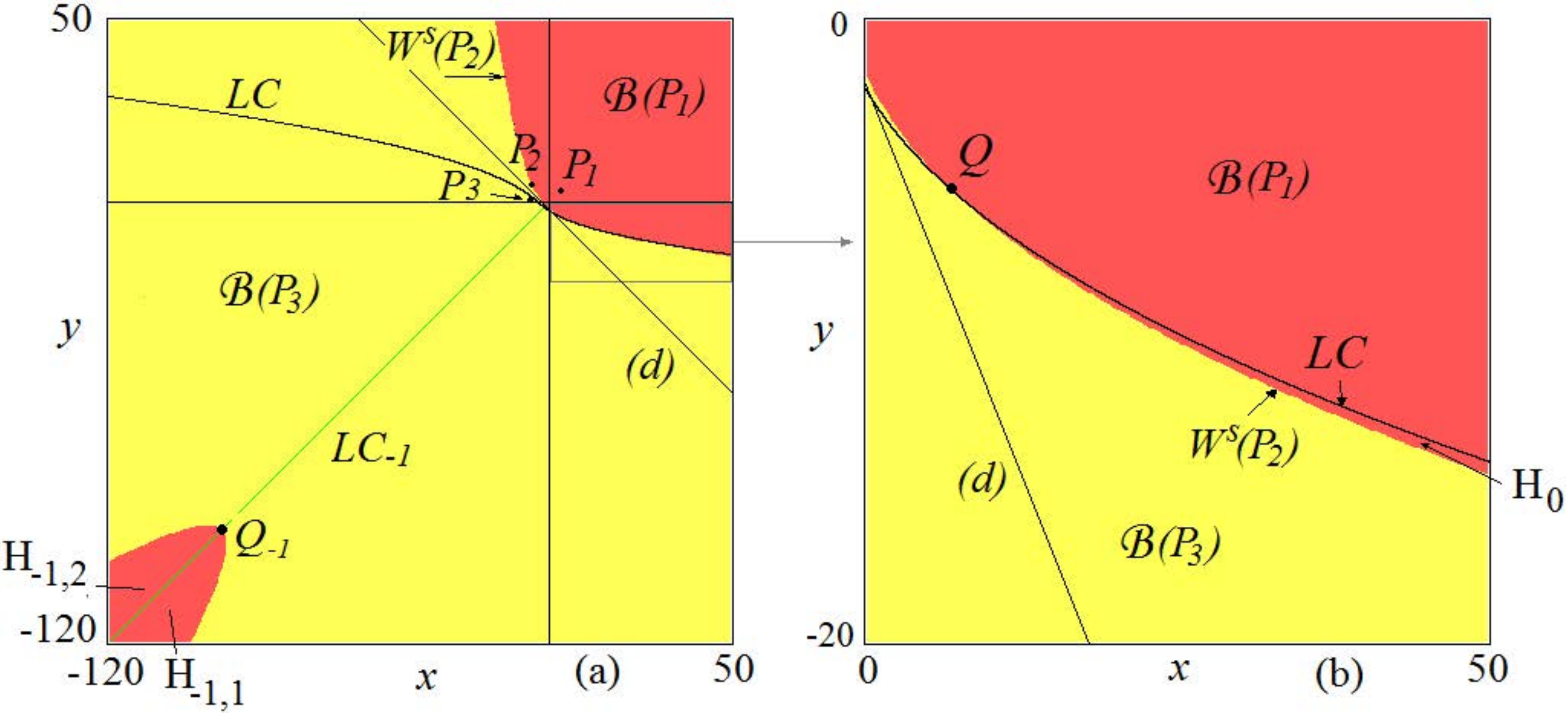}
\caption{Basins of attraction of map $G_{h}$ at $h=6$.}
\label{h6basins}
\end{center}
\end{figure}

As remarked above, for large values of $h$ the dynamics of a map $G_{h}$ is
quite regular. In fact, the map has two attracting {fixed points, }$P_{1}$ and
$P_{3}$. An example of the phase plane in such a situation is shown in
{Fig.\ref{h6basins},} for $h=6$, where $P_{1}\simeq(3.37349,3.00069),$
$P_{3}\simeq(-5.47727,0.48573)$. In that figure the basin of attraction of the
{fixed point }$P_{1}$ is shown in red, $\mathcal{B}(P_{1}),$ while that of
$P_{3}$ is shown in yellow, $\mathcal{B}(P_{3}).$ The related basins of
attraction are separated by the stable set of the saddle {fixed point }%
$P_{2}.$

The main difference with respect to the case $h=5$ is that the main branch of
the stable set of the saddle $P_{2}$ is not a critical curve and that the
basin of the {fixed point }$P_{1}$ is not connected {(for example, there are
open sets of points with both negative coordinates that have a trajectory
convergent to $P_{1}$)}. In {Fig.\ref{h6basins} it is shown the straight line
}$LC_{-1}^{j}${ up to the point }${(-1,-1)}${ and its image is the curve
}$LC^{j},${ shown in black, which intersects the line }${(d)}${ (see
(\ref{(d)})) in one point and also intersects the stable set }${W}^{{s}}%
{(}P_{2})$ in one point $Q$ (see the enlargement in {Fig.\ref{h6basins}%
})\footnote{To simplify the notation, in the figures we write only $LC$ in
place of $LC^{j}$, since these are the only arcs of interest for the dynamics
of the map, so there cannot be any confusion.}. The region denoted by $H_{0}$
in {Fig.\ref{h6basins}, which is the portion of phase plane bounded above by
the critical curve }$LC^{j}${ and below by the stable set }${W}^{{s}}{(}%
P_{2})$ on the right side of the point $Q,$ belonging to $\mathcal{B}(P_{1}%
),$\ is responsible of further preimages belonging to the basin $\mathcal{B}%
(P_{1})$.

In the Appendix, it is reported how the critical lines split the plane in
different zones named $Z_{i}$ where the sub-index $i$ denotes the number of
different rank-1 preimages. In what follows, we will use this notation. In
fact, the arc of $LC^{j}$ on the boundary of $H_{0}$ has {two merging} rank-1
preimages in the half line $LC_{-1}^{j}${ up to the point} $Q_{-1},$ while the
arc of {stable set }${W}^{{s}}{(}P_{2})$ on the right side of the point $Q$
has two distinct rank-1 preimages issuing from {the point} $Q_{-1}\in
LC_{-1}^{j},$ {one at the right side of $LC_{-1}$ (}on the boundary of the set
$H_{-1,1}${) and the other at the left side (}on the boundary of the set
$H_{-1,2}$){, forming the boundary of the set }$H_{-1}=H_{-1,1}\cup H_{-1,2}$.
Since $H_{-1}$ belongs to the zone $Z_{2}$ whose points have two distinct
rank-1 preimages, we have two more preimages of $H_{-1}$ and so on, ad
infinitum. Thus, the basin $\mathcal{B}(P_{1})$ includes also the infinite
sequence of preimages $\cup_{n\geq1}G_{h}^{-n}(H_{-1})$ which consists of
infinitely many portions of the plane (in fact, due to the properties of the
regions $Z_{k}$ described in Appendix, even if a preimage may happen to belong
to $Z_{0}$ (so having no further preimages), another one above the diagonal
exists, and thus it belongs to $Z_{k}$ with $k\geq2.$

\subsection{{Homoclinic bifurcation of $P_{2}$}}\label{SS-first-hom-bif}

{The existence of non-connected regions of the phase plane} belonging to the
basin $\mathcal{B}(P_{1})$ are relevant to prove that close to the value $h=5$
a chaotic attractor becomes a chaotic repellor which persists for
$h\in(5,\widetilde{h})$ for a certain value $\widetilde{h}$, although almost
all the points of the phase plane have a trajectory convergent to the fixed
point $P_{1}.$ Similarly, a chaotic repellor exists for $h\in(5-\varepsilon
,5)$ with $\varepsilon\gtrsim0$, and almost all the points of the plane have a
trajectory convergent to the fixed point $P_{1}.$ Thus, the occurrence of an
invariant set of positive measure (which seems chaotic) for $h=5$ as shown in
Fig.\ref{h5} is a very peculiar phenomenon in the framework of dynamical
systems theory. The peculiarity may be related to the basin $\mathcal{B}%
(P_{1})$ which is an open simply connected set only for $h=5,$ and thus
related to a peculiar behavior of the critical curves. In fact, for
$h\in(5-\varepsilon,\widetilde{h})\backslash\left\{  5\right\}  $ the basin
$\mathcal{B}(P_{1})$ is an open set but not simply connected (while for
$h\geq\widetilde{h}$ it is non connected).

We show that there exists a particular value $\widetilde{h}\simeq5.032$ such
that for $h>\widetilde{h}$, there exists an invariant unbounded {absorbing
region $\mathcal{A}$},{ which is bounded by segments of images of finite rank
of a \emph{generating segment} $g$ }belonging to the critical curve
$LC_{-1}^{j}$ for $x<-1$ , that is, below the line $(d)$ of vanishing
denominator. For $h<\widetilde{h}$ such an invariant region does not
exist\footnote{For $h=5$ there exist an invariant area, but it is not
absorbing.}. Moreover, for $h>\widetilde{h}$ (resp. $h<\widetilde{h}$) the
saddle fixed point $P_{2}$ is not homoclinic (resp. is homoclinic).

\subsubsection{Situation before the contact and qualitative description of the
contact bifurcation}

\label{SS241}

For $h>\widetilde{h}$ (when the unbounded absorbing region $\mathcal{A}$
exists) the stable set of the saddle separates two basins of attraction: the
basin $\mathcal{B}(P_{1})$ of the attracting fixed point $P_{1}$ and the basin
$\mathcal{B}(\mathcal{A})$ of the absorbing area $\mathcal{A}$ (inside which
several attracting sets may exist). That is, the frontier of these two basins
is that stable set: ${W}^{{s}}{(}P_{2})=\partial\mathcal{B}(P_{1}%
)=\partial\mathcal{B}(\mathcal{A}).$
\begin{figure}
[h]
\begin{center}
\includegraphics[scale=0.42]{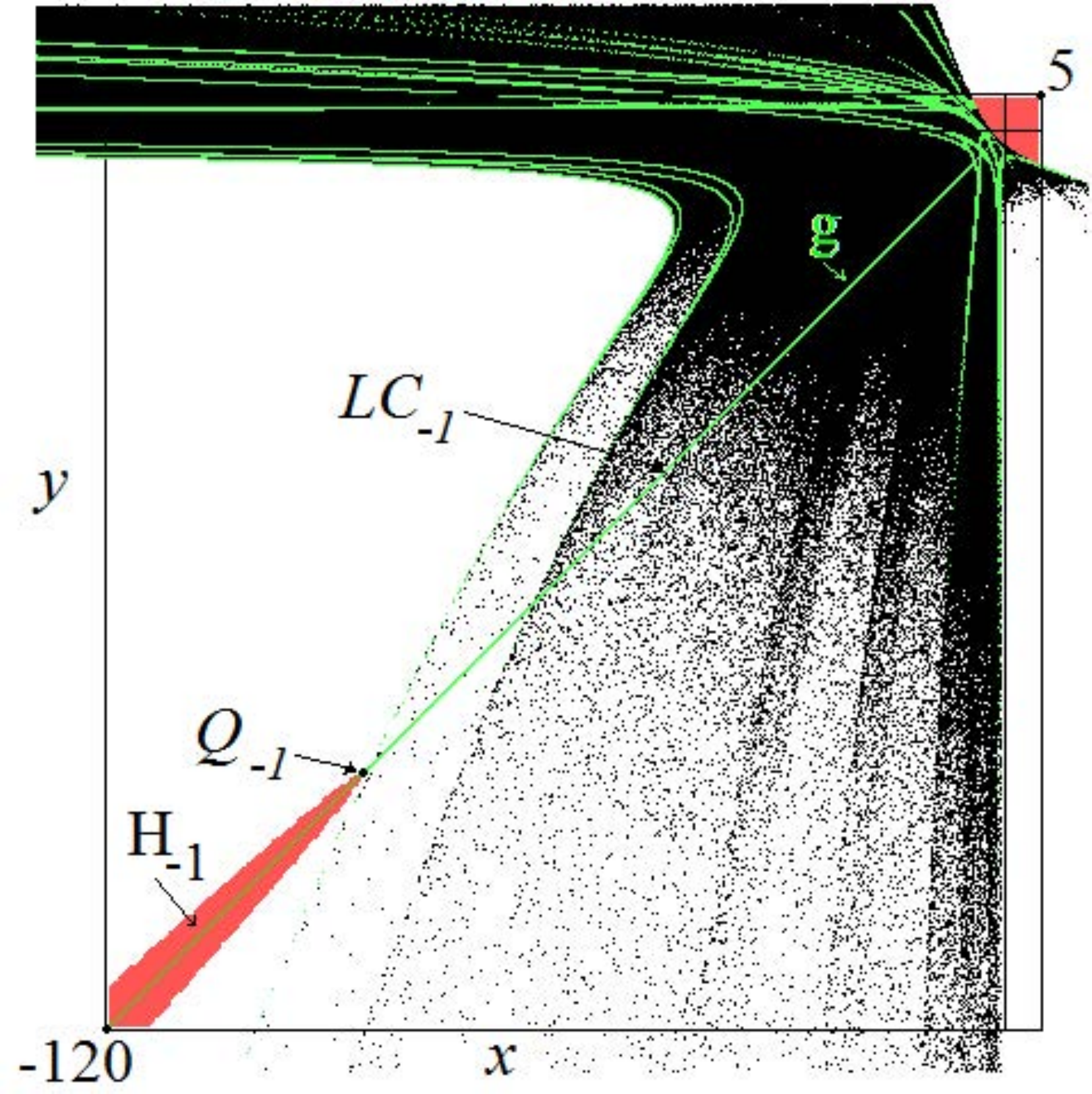}
\caption{Phase portrait of map $G_{5.032}$. The white area belongs to the
basin of the invariant chaotic area. The red area belongs to the basin
$\mathcal{B}(P_{1})$: In green, the images of the generating segment $g$ of
the critical line $LC_{-1}^{j}$. The region $H_{-1}$, which is part of the
basin $\mathcal{B}(P_{1})$, is close to contact with the chaotic area bounded
the arcs given by the images of the generating segment.}
\label{lastchaos}
\end{center}
\end{figure}

In {Fig.\ref{lastchaos}} we illustrate the unbounded chaotic area at
$h=5.032$, whose boundary consists of a finite number of images of the
generating segment $g$ which is the arc of $LC_{-1}^{j}$ belonging to the
invariant area itself (see \cite{MGBC 96}). It can be seen that the boundary
is very close to contact the point $Q_{-1}\in LC_{-1}^{j}$ on the boundary of
the set $H_{-1}$ which belongs to the basin $\mathcal{B}(P_{1}).$ Recall that
this boundary belongs to the stable set of the saddle, ${W}^{{s}}{(}P_{2}),$
while the left branch of the unstable set of the saddle, ${W}^{{u,l}}{(}%
P_{2}),$ entering the chaotic area, is dense in that area (differently, the
other branch ${W}^{{u,r}}{(}P_{2})$ issuing from the saddle $P_{2}$ is a
branch connecting the saddle with the attracting fixed point $P_{1})$. When
such a contact occurs, say at $h=\widetilde{h},$ that is, when the boundary
of\ $H_{-1}$ has a contact with the critical curve bounding the chaotic area,
then this contact leads (for $h<\widetilde{h}$) to the destruction of the
invariant chaotic area and simultaneously to the appearance of homoclinic
points of the saddle $P_{2}$ (but on the left side only).
\begin{figure}
[ptb]
\begin{center}
\includegraphics[scale=0.3]{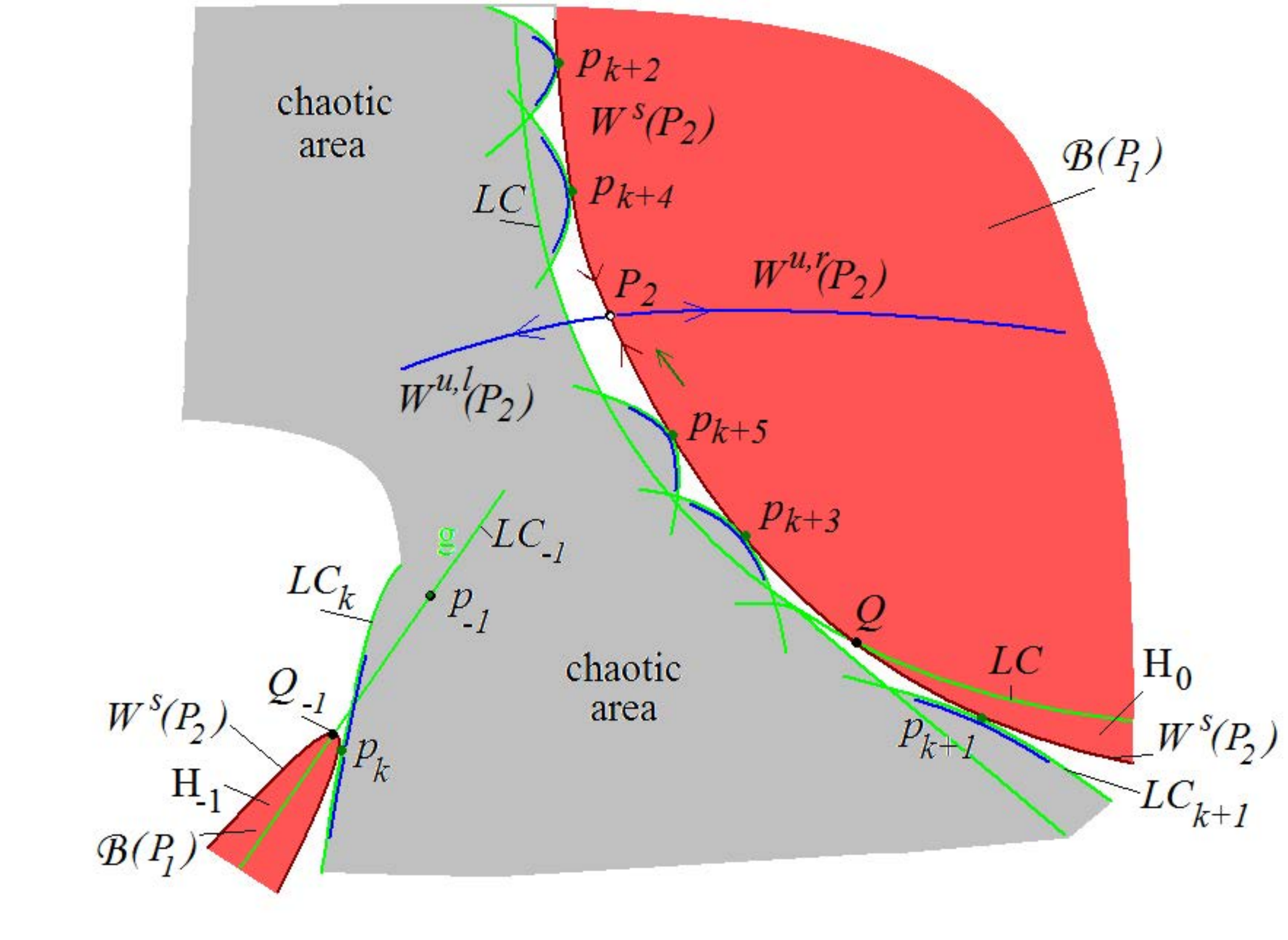}
\caption{Qualitative description of the contact bifurcation occurring at
$h=\widetilde{h}.$}
\label{HomBifSaddle}
\end{center}
\end{figure}
The role played by contact bifurcations of this kind have been described in
several works (see e.g. \cite{FournierEtAl} and \cite{MGBC 96}), see also the
bifurcations called \textquotedblleft crisis\textquotedblright\ in
\cite{Grebogi83,Grebogi87}).

The contact bifurcation occurring at $h=\widetilde{h}$ is qualitatively shown
in {Fig.\ref{HomBifSaddle}}. The contact point between the region $H_{-1}$
(whose boundary belongs to the stable set ${W}^{{s}}{(}P_{2}))$ and the
chaotic area bounded by arcs of critical curves, is denoted by $p_{k}\in
LC_{k}^{j}$ for a suitable integer $k$, and it is the image of a point
$p_{-1}\in g\subset LC_{-1}^{j}.$ If the area is truly chaotic then the
closure of the unstable set ${W}^{{u,l}}{(}P_{2})$ is exactly the chaotic area
(i.e. ${W}^{{u,l}}{(}P_{2})$ is dense in the area), which means that the arcs
of critical curves are limit sets of the unstable set (arcs shown in blue in
{Fig.\ref{HomBifSaddle}}). The image of $p_{k}$ is the point $p_{k+1}\in
LC_{k+1}^{j}$ belonging to the boundary of $H_{0}$ on the arc of the stable
set ${W}^{{s}}{(}P_{2})$, so that the same property holds for the forward
images which are converging to the saddle fixed point $P_{2}$: that is, they
are contact points between critical arcs on the boundary of the invariant area
$\mathcal{A}$ and the main branch of the stable set ${W}^{{s}}{(}P_{2})$.

\subsubsection{Situation after the contact bifurcation}

For $h<\widetilde{h}$ when the critical curves intersect the regions\ $H_{-1}$
and $H_{0}$ also the unstable set ${W}^{{u,l}}{(}P_{2})$ is crossing these
regions, and thus is crossing the stable set ${W}^{s}{(}P_{2})$ leading to
homoclinic points of the saddle, and since an invariant absorbing area no
longer exists the generic trajectory converges to the attracting fixed point
$P_{1}$.

Recall that for noninvertible maps we cannot refer to a general theorem
stating that when a homoclinic orbit of a saddle cycle exists then close to
the homoclinic orbit we can find an invariant set which is chaotic (in the
sense of Devaney \cite{Devaney} and of Li and Yorke \cite{Li}), because the
standard theorem holds for diffeomorphisms. However, also for endomorphisms
(having not a unique inverse) it is possible to prove that following the
homoclinic orbit in the proper way, with the suitable inverses applied to the
homoclinic points, it is possible to show that a horseshoe can be rigorously
defined, leading to the result. Examples can be found in \cite{Gard96} and
\cite{GarSus98}.

Thus, for $5<h<\widetilde{h}$ the chaotic attractor becomes a chaotic
repellor, the only attracting set is the fixed point $P_{1},$ and its basin
$\mathcal{B}(P_{1})$ is almost the whole plane. The points of the plane on the
left side of the main branch of the stable set of the saddle $P_{2}$ that are
leaving that side converging to the attracting fixed point $P_{1}$ are doing
so ultimately crossing through the regions called $H_{-1}$ and $H_{0}$, where
$H_{0}$ is always defined as above:\ it is the strip between the main branch
of the stable set of $P_{2}$ and the arc of critical curve $LC^{j}$ above it,
which exists as long as $h>5$ (as shown below). Notice that also the unstable
set of the saddle $P_{2}$ for $5<h<\widetilde{h}$ has points crossing through
these regions. That is: arcs belonging to the unstable set ${W}^{{u,l}}%
{(}P_{2})$ as well as arcs belonging to the critical lines $LC_{n}^{j}$ for
some $n>1$ have points which cross through $H_{-1}$ and $H_{0}$ and then
converge to the attracting fixed point $P_{1}.$

\subsection{Final bifurcation for $h=5$. Role of the critical lines}\label{SSfinal}

{At $h=5$, the scene described in Section \ref{SS-first-hom-bif} is no longer
possible because the strip $H_{0}$ reduces to the arc $LC^{j}$ which merges
with the main branch of the stable set of the saddle $P_{2}$, and thus the
portion $H_{-1}$ reduces to an arc of $LC_{-1}^{j}.$ The portion of plane
bounded by the arc of curve $LC^{j}$ (belonging to the resolvent curve given
in Eq.(\ref{resolvant})) is invariant: it is the situation shown in
Fig.\ref{h5}. Notice that the saddle\ $P_{2}$ is homoclinic, since the whole
segment $LC_{-1}^{j}$ and all its preimages belong to the stable set of
$P_{2}$ and intersections between the stable and unstable sets are infinitely
many, but only on the left side of the arc $LC^{j}.$ This means that,
differently from the case for $5<h<\widetilde{h}$, even if the saddle\ $P_{2}$
is homoclinic\ no one point belonging to the unstable set ${W}^{{u,l}}{(}%
P_{2})$ can have a trajectory convergent to $P_{1}$ (since no one point can be
mapped to the right side of $LC^{j}$) and similarly {no point on the images of
the critical line $LC_{-1}^{j}$}, of any rank, can converge to $P_{1}.$ }

So we have seen that for $5\leq h<\widetilde{h}$ the saddle $P_{2}$ is
homoclinic only on the left side.
\begin{figure}
[h]
\begin{center}
\includegraphics[scale=0.35]{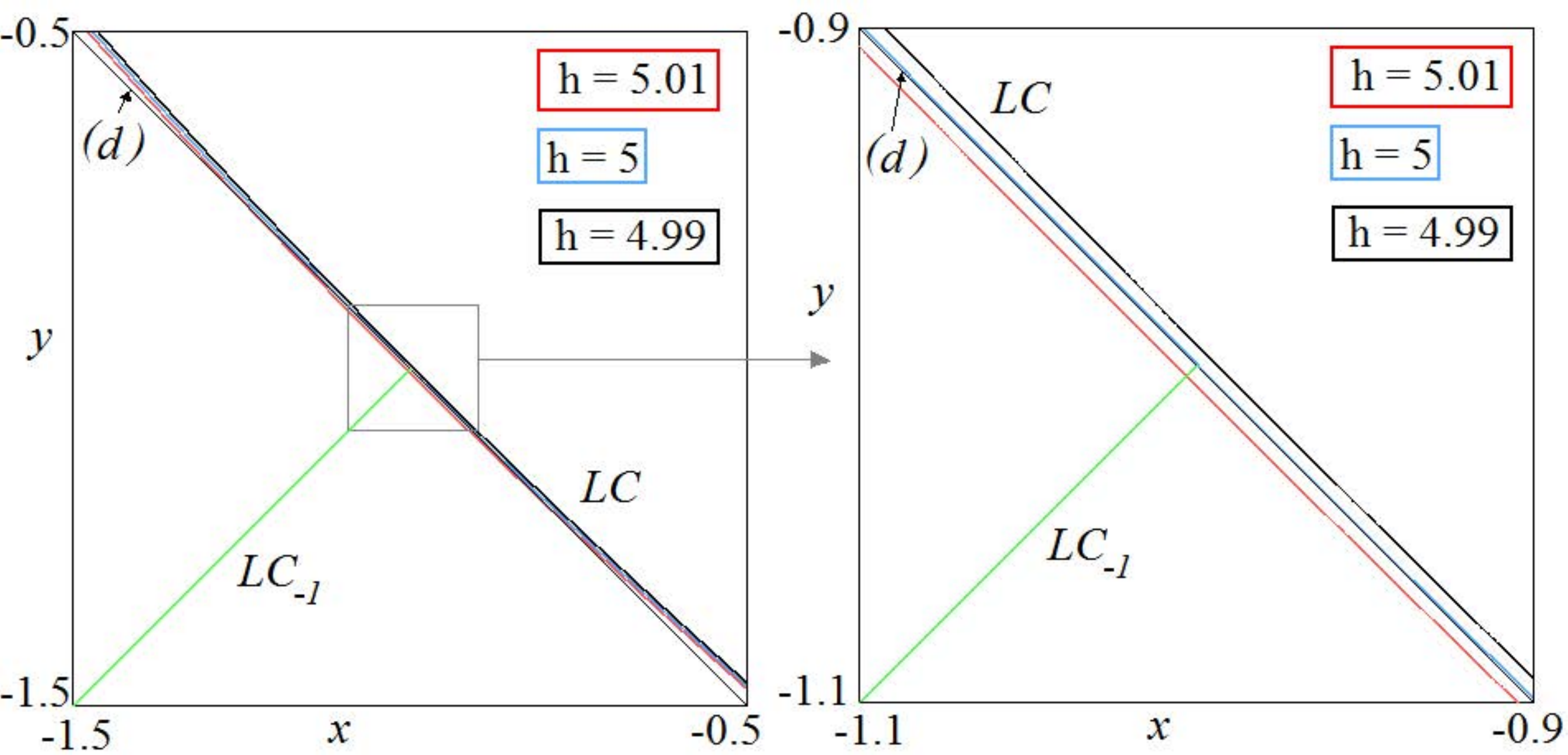}
\caption{Images of an arc of $LC_{-1}^{j}$ for $x\leq-1$ at $h=5.01$, $h=5$
and $h=4.99$.}
\label{Bifh5}
\end{center}
\end{figure}
The characteristic commented above associated with the Boros-Moll map at $h=5$
may be related to the critical curves as follows. For $5<h<\widetilde{h}$ the
image of the critical curve $LC_{-1}^{j}$ for $x\leq-1$ is an unbounded branch
$LC^{j}$ which intersects the line of vanishing denominator $(d)$ in two
points, so that $LC_{1}^{j}$ consists in three unbounded components. In
{Fig.\ref{Bifh5}} we show a portion of phase plane, an enlargement close to
$(-1,-1)$ at $h=5.01$, at this value of the parameter almost all the points
have a trajectory converging to $P_{1}\simeq(3.00375,$ $3.00000).$ At $h=5$
the image of $LC_{-1}^{j}$ for $x\leq-1$ is the branch $LC^{j}$ which merges
with the main arc of the stable set of $P_{2},$ which is tangent (from above)
to $(d)$ in the point $(-1,-1)$, and it is invariant, so that $LC^{j}%
=LC_{n}^{j}$ for any $n\geq1.$ For $h<5$ the image of $LC_{-1}^{j}$ for
$x\leq-1$ is the branch $LC^{j}$ completely above $(d)$, and thus its points
converge to the attracting fixed point $P_{1}.$ {Clearly, all the points in
the strip {$H_{0}$} between the main arc of the stable set of $P_{2}$ and the
arc $LC^{j}$ are also converging to the attracting fixed point $P_{1}$,} and
so do the points of its preimage belonging to a strip including $LC_{-1}^{j}$
as well all the further preimages of any rank. For instance, at $h=4.99$, a
chaotic repellor still exists (since the fixed points $P_{2}$ and $P_{3}$\ are
homoclinic, as well as infinitely many repelling cycles exist and are
homoclinic), and almost all the points are converging to $P_{1}\simeq
(2.99625,$ $3.00000).$
\begin{figure}
[h]
\begin{center}
\includegraphics[scale=0.37]{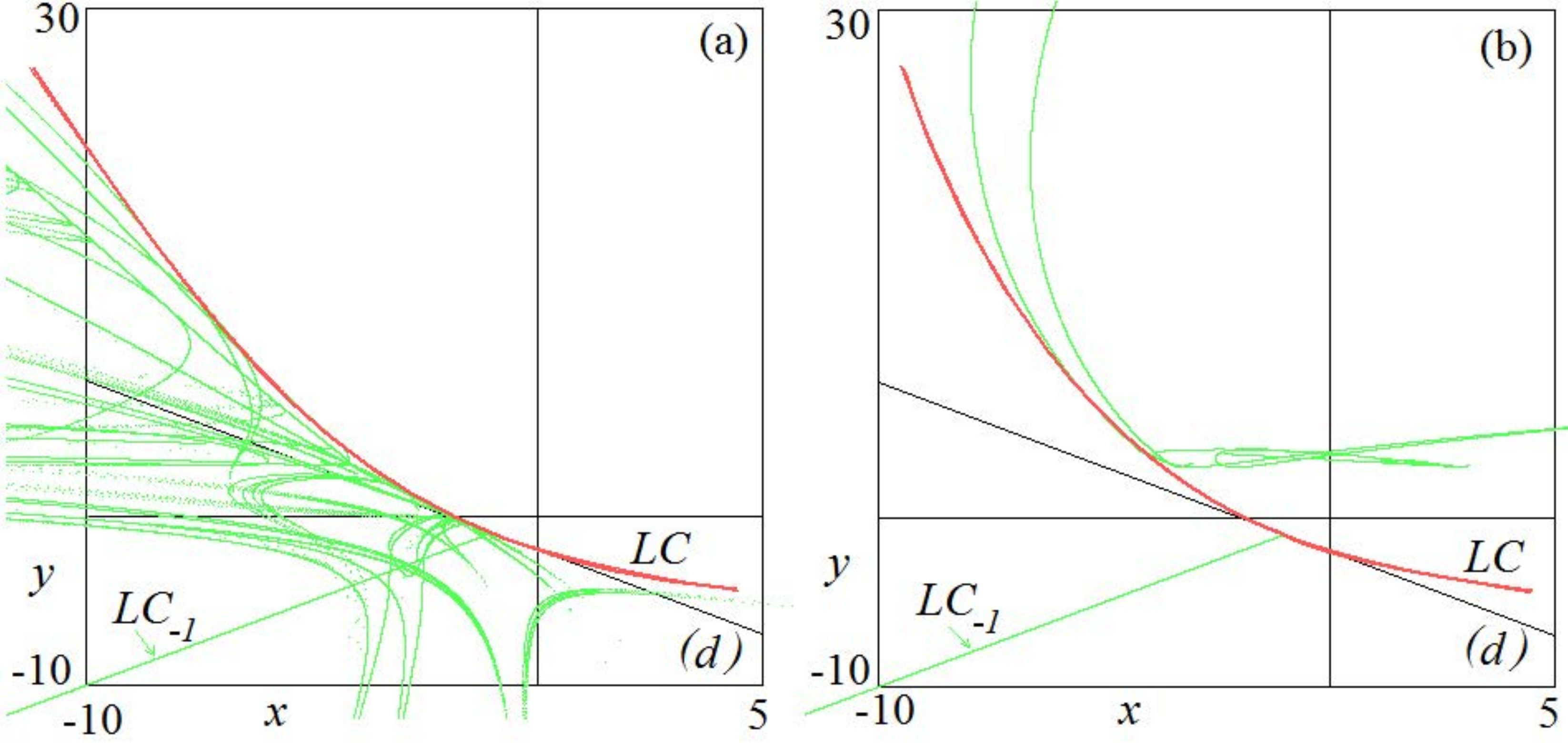}
\caption{Critical curves $LC_{k}^{j}$ for $1\leq k\leq4$ at $h=5.01$ in (a),
at $h=4.99$ in (b).}
\label{archiLCi}
\end{center}
\end{figure}

For $h<5$ not only the critical curves are on the right side of the main
branch of the stable set of $P_{2}$, but also the unstable set ${W}^{{u,l}}%
{(}P_{2})$ has again arcs going to the right side and converging to $P_{1}.$

In {Fig.\ref{archiLCi}} we illustrate an arc of $LC_{-1}^{j}$ for $x\leq-1,$
$LC^{j}$ in red and its first images, at $h=5.01$, which are all on the left
side of the curve $LC^{j},$ while at $h=4.99$ the images are all on the right
side to the $LC^{j}$ (clearly for $h=5$ this branch of $LC^{j}$ is invariant).

{The above study evidences that the arc $LC^{j}$ is intersecting the main arc
of the {stable set } ${W}^{{s}}{(}P_{2})$ in one point $Q$ as long as $h>5$,
they are merging when $h=5$, and $LC^{j}$ is above that stable set for $h<5.$}

\section{Route to chaos}\label{Sroute}

\begin{figure}
[h]
\begin{center}
\includegraphics[scale=0.37]{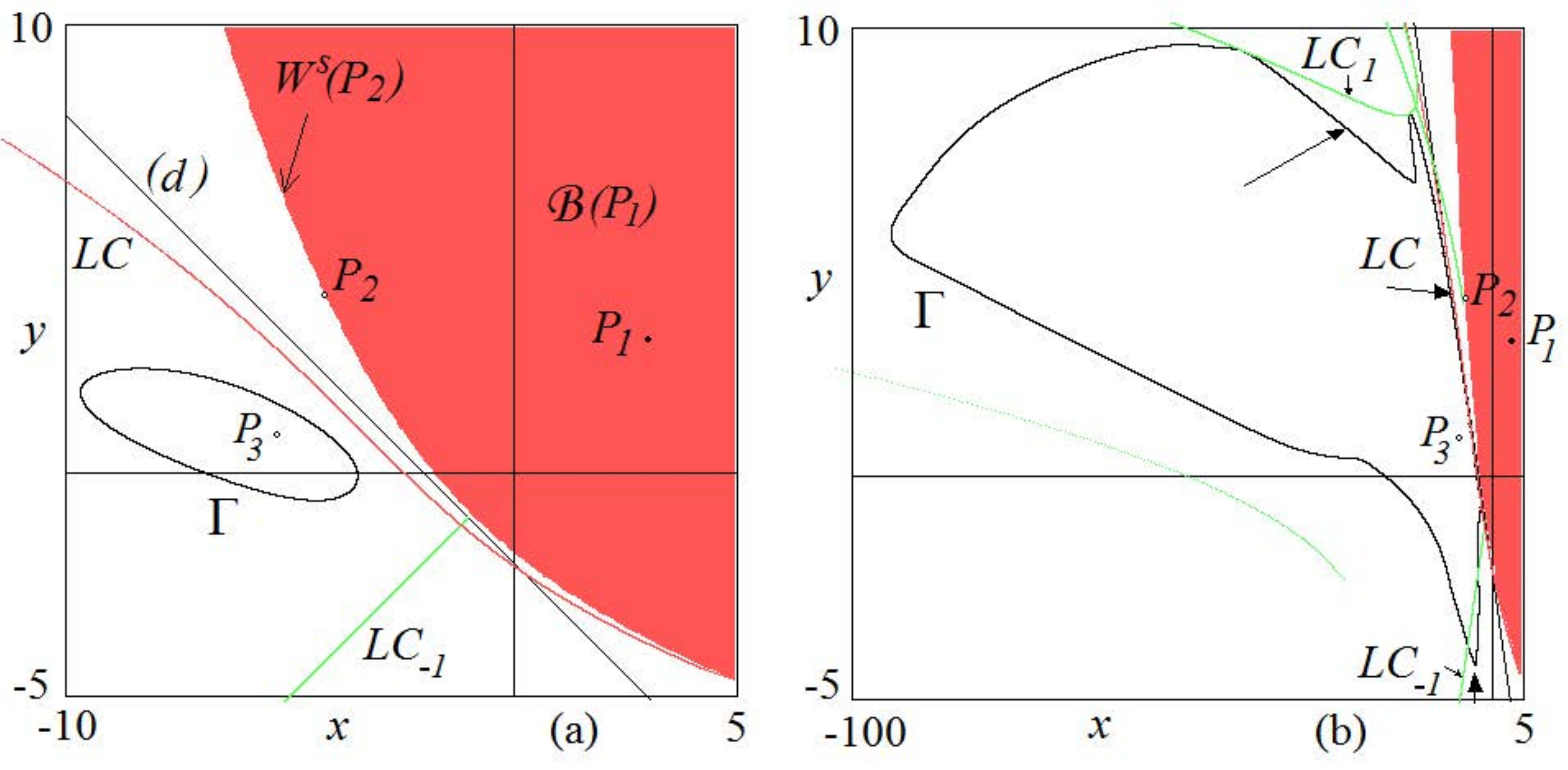}
\caption{Bifurcation of the closed curve $\Gamma$. In (a) $h=5.6,$ smooth
curve not intersecting $LC_{-1}^{j}.$ In (b) $h=5.44,$ $\Gamma$ intersects
$LC_{-1}^{j}.$ }
\label{GamBif}
\end{center}
\end{figure}

In order to describe the dynamics of map $G_{h}$ (leading to the unbounded
chaotic set shown in {Fig.\ref{lastchaos}}) we account some of the phenomena
that appear when the parameter $h$ is decreased to $h=5$ after the $NS$
bifurcation of the fixed point $P_{3},$ that is, decreasing $h$ from
$h_{NS}\simeq5.6105.$ The $NS$ bifurcation is of supercritical type, as can be
seen from the attracting closed curve $\Gamma$ which appears surrounding the
repelling focus $P_{3}$, an example is illustrated in {Fig.\ref{GamBif}a}. The
restriction of the map to the invariant curve in general leads either to an
attracting cycle (when the rotation number is rational and the closed curve
consists of a saddle-node connection) or to quasiperiodic orbits, dense in the
closed curve (when the rotation number is irrational). It is worth noting that
as long as the curve $\Gamma$ is in a region without intersections with the
critical curve $LC_{-1}^{j}$ then all the points inside the area bounded by
$\Gamma$ cannot be mapped outside the area (i.e. the area bounded by $\Gamma$
is invariant). At the same time, the points external to that area which are
attracted from $\Gamma$ have the trajectory completely outside $\Gamma$ (i.e.
the points are approaching the closed curve from outside). The dynamics in the
phase plane change, as well as the global shape of the curve $\Gamma,$ after a
crossing of the critical curve $LC_{-1}^{j}.$

\begin{figure}
[h]
\begin{center}
\includegraphics[scale=0.37]{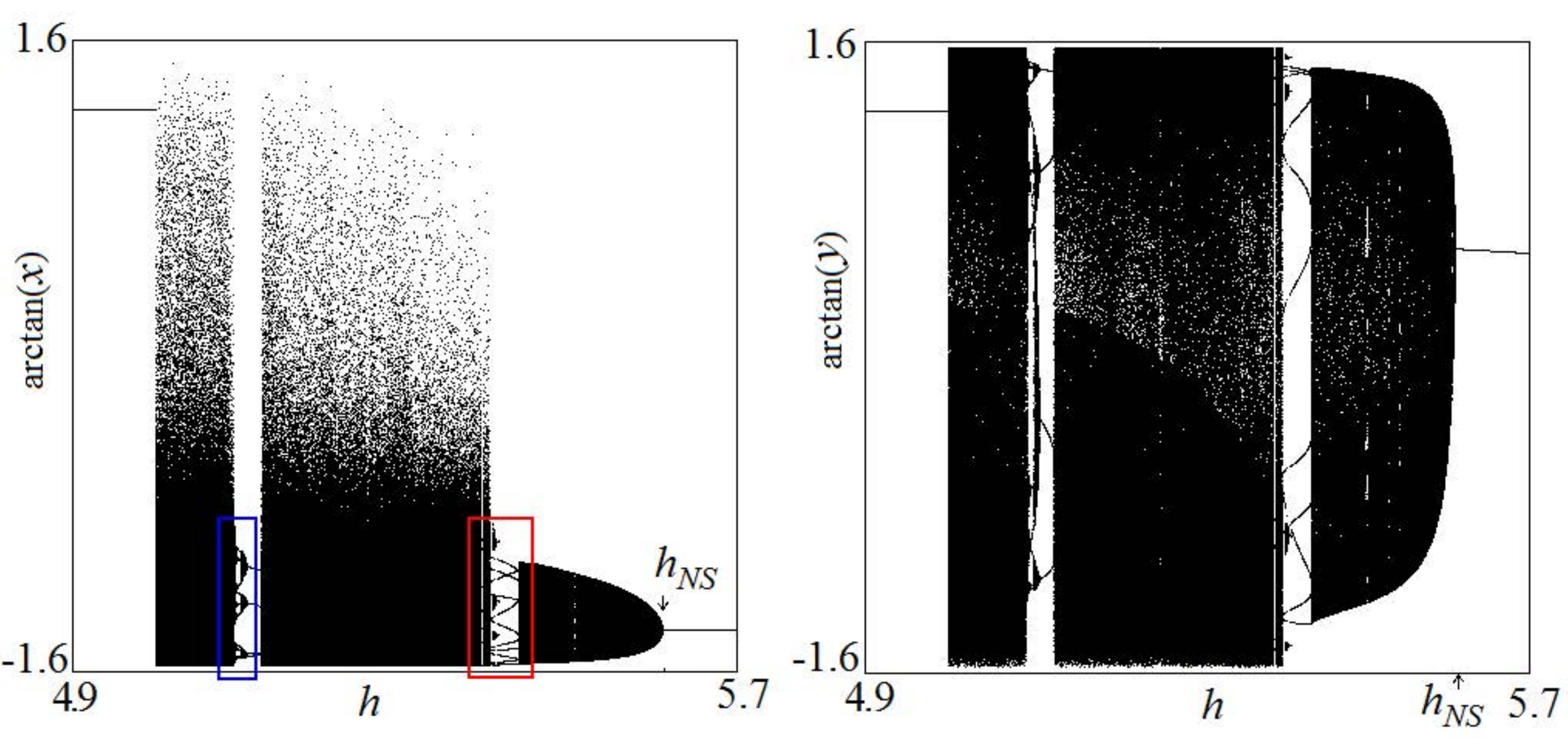}
\caption{One-dimensional bifurcations diagrams.}
\label{1DA}
\end{center}
\end{figure}
\begin{figure}
[h]
\begin{center}
\includegraphics[scale=0.37]{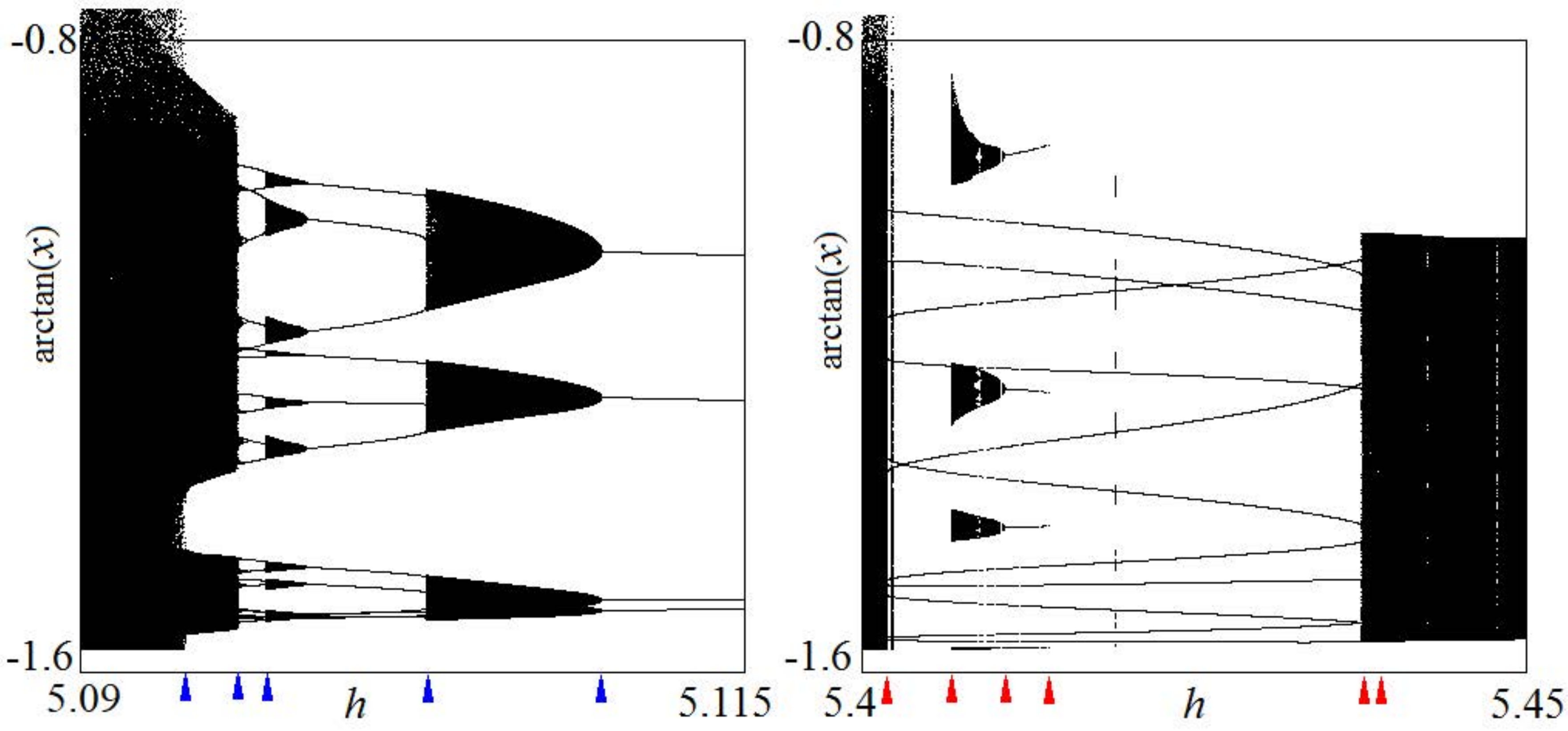}
\caption{One-dimensional bifurcations diagrams. Enlargements of the rectangles
shown in color in Fig.\ref{1DA}a.}
\label{1DB}
\end{center}
\end{figure}
Decreasing $h$ a contact between $\Gamma$ and $LC_{-1}^{j}$ occurs at
$h\simeq5.506$, after which there is a portion of $\Gamma$ which is below the
critical curve and thus the portion of area\ bounded by $\Gamma$ and crossing
$LC_{-1}^{j}$ is folded on $LC^{j}$ so that $\Gamma$ becomes tangent to
$LC^{j}$ in two points (the rank-1 images of the two intersection points
$\Gamma\cap LC_{-1}^{j}$). And clearly the contacts between $\Gamma$ and the
critical curves $LC_{n}^{j}$ persist for any $n$. In {Fig.\ref{GamBif}b} it is
well evident the tangency between $\Gamma$ and the critical curve $LC_{1}%
^{j}.$

This mechanism is a peculiar phenomenon related to noninvertibility (i.e. it
cannot occur in maps with a unique inverse). It is related to the folding of
the phase plane due to the critical curves, and breaks the invariance of the
area bounded by the closed curve $\Gamma.$ In fact, all the points in the area
bounded by $\Gamma$ and $LC_{-1}^{j}$ are mapped outside $\Gamma,$ between
$\Gamma$ and $LC^{j}$, {and its further images between $\Gamma$ and
$LC_{1}^{j}$, and so on (see the related properties in {\cite{FrouzakisEtAl}
and \cite{MGBC 96}}). In particular, this fact allows the existence of cycles
of the map with periodic points both inside and outside the area bounded by
the closed curve.} We shall see an example in the next subsection.

\medskip

Let us first illustrate a sequence of attractors and their bifurcations via
one-dimensional bifurcation diagrams as a function of the parameter $h$. Since
we know that unbounded attracting sets exist, in Fig.\ref{1DA} we show the
scaled variable (as we have done in Fig.\ref{h5}b) to better emphasize when an
attractor is unbounded. In Fig.\ref{1DA} it is evident the range in which
decreasing $h$ from $h_{NS}$ the closed invariant curve $\Gamma$ increases in
size, since approaching the straight line (d) the shape becomes wider. In the
same figure we see that there are two large windows associated with cycles and
interesting dynamics, which are enlarged in Fig.\ref{1DB} and commented in the
next subsections.

\subsection{Bifurcations occurring in the first enlargement.}

\begin{figure}
[h]
\begin{center}
\includegraphics[scale=0.37]
{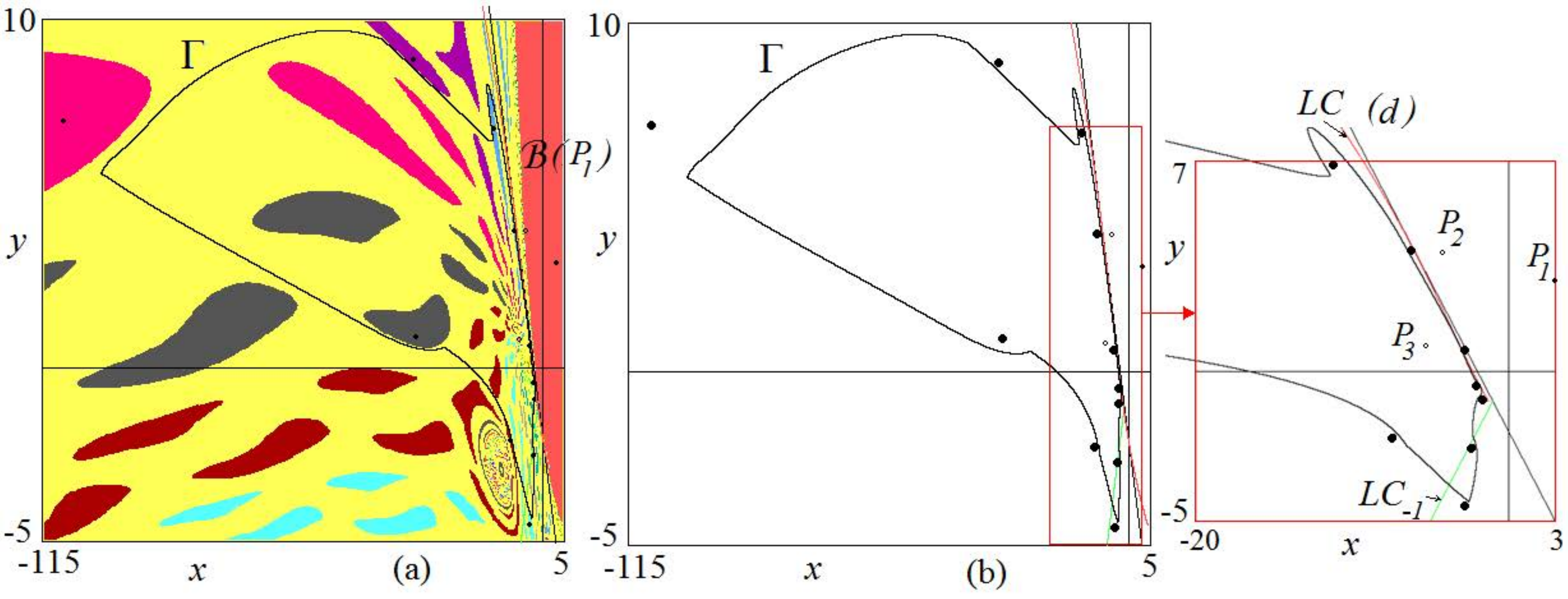}
\caption{Coexistence of the closed curve $\Gamma$ and an attracting 11-cycle
at {$h=$}$5.4378.$}
\label{coexistence11Gam}
\end{center}
\end{figure}

As noticed above, the oscillations on the closed curve $\Gamma$ become
pronounced as $h$ is decreased. Moreover, other attracting sets appear by
saddle-node bifurcation. An example is shown in {Fig.\ref{coexistence11Gam}.
At }${h=}5.4379$ besides $P_{1}$ only the closed curve $\Gamma$ is attracting,
while at {$h=$}$5.4378$ we can see that a pair of 11-cycles have appeared.
{One of them is attracting, while the other is a 11-cycle saddle whose stable
set bounds the basin of attraction of the attracting 11-cycle, separating its
basin from the basin of $\Gamma$.} In {Fig.\ref{coexistence11Gam}a we show in
red (resp. yellow) the basin of }$P_{1}$ (resp. the closed curve $\Gamma${)
and we show with different colors the basins of the 11 fixed points of the
11-th iterate of the map }$G_{h}${, to emphasize the shape of the basin of the
attracting 11-cycle. In} {Fig.\ref{coexistence11Gam}b and related enlargement
it can be seen that some points of the attracting 11-cycle are outside }the
closed curve $\Gamma${ and part are inside. One more peculiarity related to
the }immediate basins of the 11 fixed points\ of map $G_{h}^{11}$ is that the
stable set of the 11 saddles (related to the 11-cycle saddle) leads to a
closed curve, which is possible only in noninvertible maps, due to the
crossing of the critical curves of the map (other examples and explanation of
the mechanism can be found in {\cite{MGBC 96}}).
\begin{figure}
[h]
\begin{center}
\includegraphics[scale=0.37]{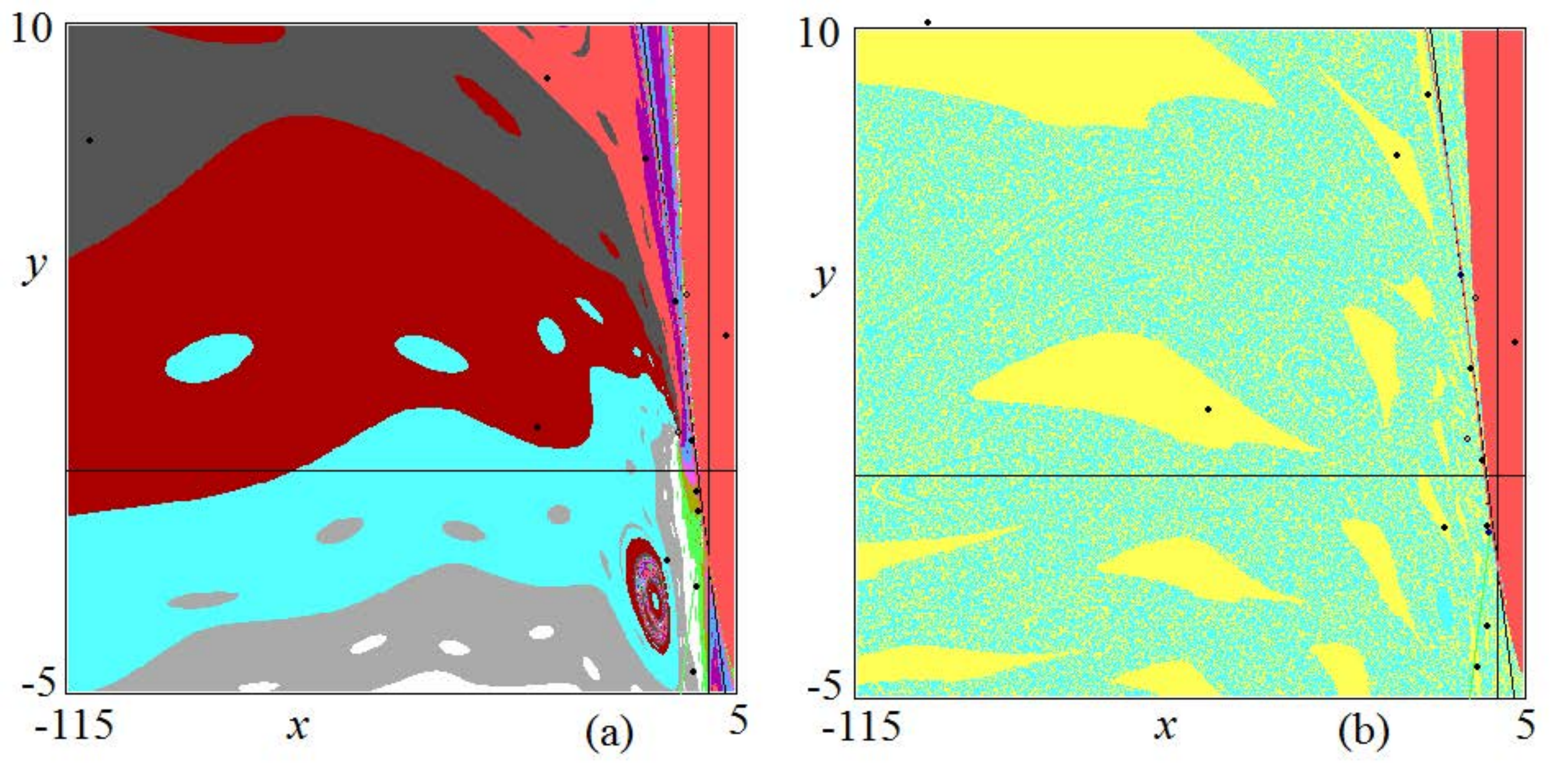}
\caption{In (a) {$h=$}$5.437,$ attracting 11-cycle and its basin of attraction
for the 11-th iterate of the map. In (b) {$h=$}$5.412,$ coexistence between
the attracting 11-cycle and the attracting 4-cycle (in yellow and azure their
respective basin of attraction).}
\label{Double}
\end{center}
\end{figure}
From the same {Fig.\ref{coexistence11Gam}} we can see that the curve $\Gamma$
is quite close to the stable set of the saddle 11-cycle on the boundary of the
basin of attraction. A contact between these two invariant sets may lead to
the disappearance of the closed curve $\Gamma,$ and its transition to some
repelling set. In any case, at {$h=$}$5.437$ {besides }$P_{1}$ only the
11-cycle is attracting, and the stable set of the saddle 11-cycle has now a
quite different shape, confirming the disappearance of the closed invariant
curve, as shown in {Fig.\ref{Double}a, where the crossing of }$LC^{j}${ of the
stable set of the saddle leads to the disconnected components evidenced in
Fig.\ref{Double}a for the 11-th iterate of the map.}

{Decreasing $h$, one more saddle-node bifurcation occurs, now giving rise to a
4-cycle. As can be seen in Fig.\ref{1DB}b this occurs at $h\simeq5.4125$}
leading to another coexistence of attracting sets on the left side of the main
branch of the stable set of $P_{2}$. {In Fig.\ref{Double}b }at {$h=$}$5.412$
{besides }$P_{1}$ (with basin in red) the basins of attraction of the 11-cycle
(in yellow) and of the 4-cycle (in azure) are shown. The basins' structure of
the two coexisting attractors is clearly fractal, which means that the saddle
11-cycle on the immediate basin of the 11-cycle is now homoclinic on one side
(that external to the immediate basin which is clearly visible in yellow via
bounded closed pieces).
\begin{figure}
[h]
\begin{center}
\includegraphics[scale=0.37]{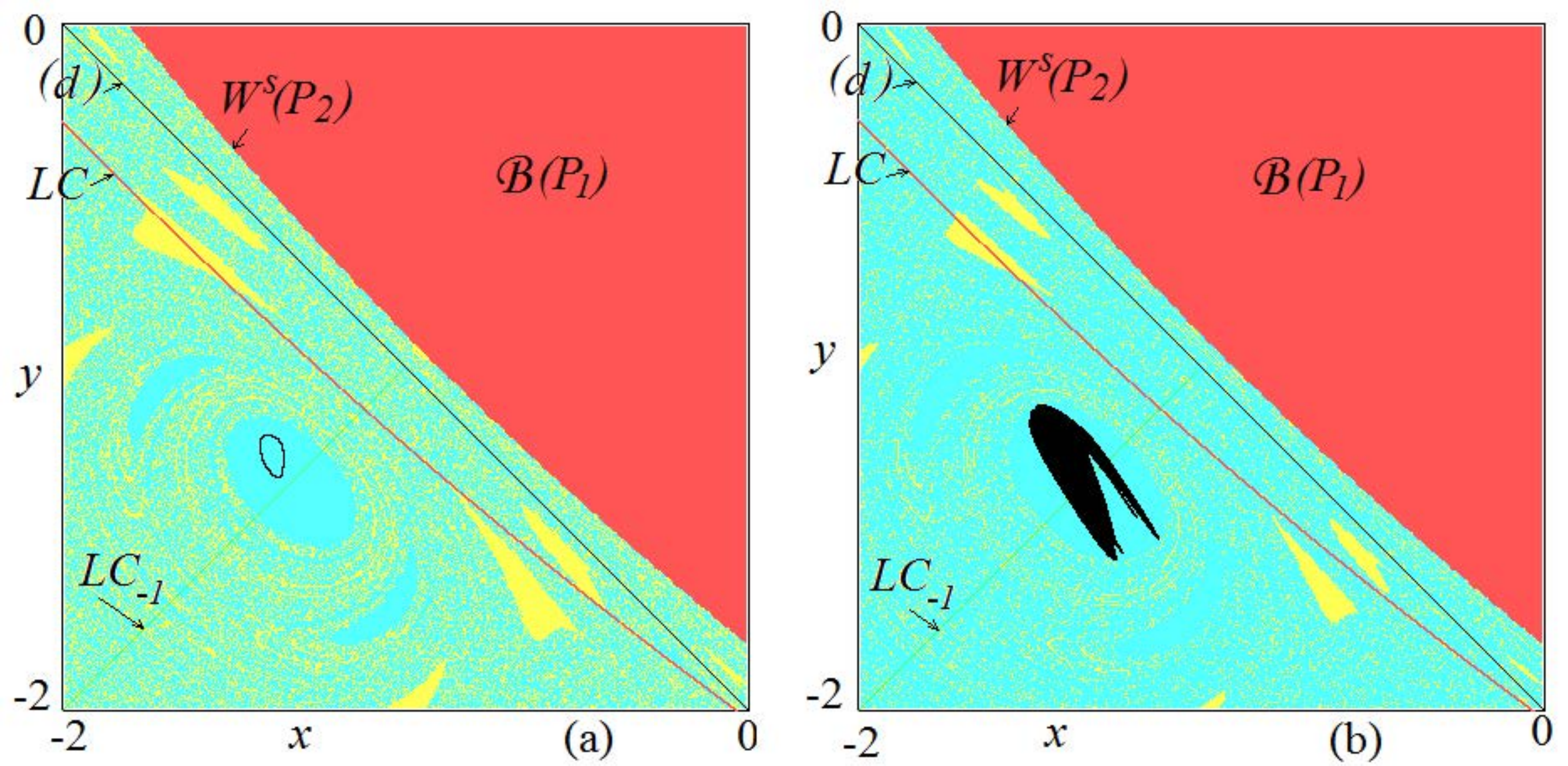}
\caption{In (a) $h=5.41,$ $NS$ bifurcation of the 4-cycle. In (b) $h=5.40676,$
chaotic piece showing that the 4-cycle is a snap-back repellor.}
\label{sbr}
\end{center}
\end{figure}
As evidenced in Fig.\ref{1DB}b the attracting 4-cycle undergoes a
supercritical $NS$ bifurcation at $h\simeq5.411$ which leads to 4-cyclic
closed invariant curves (see an example in the enlargement of Fig.\ref{sbr}a
at $h=5.41$). The closed curves undergo bifurcations to annular chaotic
regions and to 4-cyclic chaotic pieces (see the portion of phase plane in
Fig.\ref{sbr}b at $h=5.40676$). This sequence, as described in \cite{Gard94}
(see also \cite{MGBC 96}), leads to the homoclinic bifurcation of the
expanding 4-cycle, repelling focus, which becomes a snap-back repellor
(following Marotto \cite{Marotto}). The chaotic piece of the 4-cyclic set
shown in Fig.\ref{sbr}b is very close to the boundary of its basin of
attraction, and a contact with the basin leads to the disappearance of the
4-cyclic chaotic pieces, which become a chaotic repellor. Already at
$h=5.40675$ a chaotic transient can be observed and the only attractor left on
the left side of the main arc of the stable set ${W}^{s}{(}P_{2})$ is the
11-cycle (with a chaotic repellor on its basin boundary).%
\begin{figure}
[h]
\begin{center}
\includegraphics[scale=0.32]{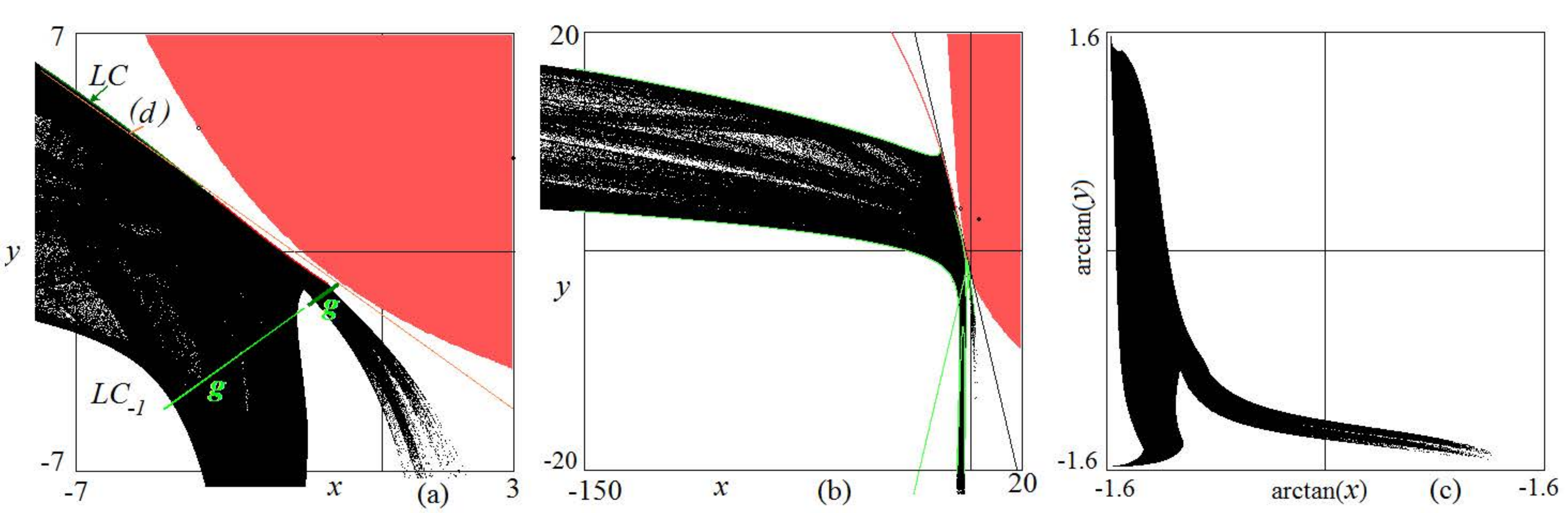}
\caption{Unbounded absorbing area, with chaotic dynamics, at $h=5.4.$ In (a)
the generating segments on $LC_{-1}^{j}$ are evidenced. In (b) the attracting
set is shown in a larger portion of the phase plane. In (c) the attracting set
is shown the plane scaled as $(\arctan(x),\arctan(y)).$}
\label{Unbounded3}
\end{center}
\end{figure}

The attracting 11-cycle disappears with the companion saddle cycle in a
(reversed) saddle-node bifurcation. After this saddle-node bifurcation we
cannot observe a simple attracting cycle, but a closed invariant absorbing
area exists, inside which the dynamics seem chaotic. At $h=5.4$ the area shown
in Fig{.\ref{Unbounded3}a intersects the critical line }$LC_{-1}^{j}${ in two
segments, forming together the generating segments whose images bound the
absorbing area via segments of critical curves. The area is unbounded as in
fact the image of the shortest segment (evidenced in dark green in
Fig.\ref{Unbounded3}a) is mapped in an arc of }$LC^{j}${ which crosses the
line (d) of vanishing denominator (evidenced with a dark green arrow in
Fig.\ref{Unbounded3}a), thus its image consists in two unbounded arcs of
}$LC_{1}^{j}${ on the boundary of the area. In Fig.\ref{Unbounded3}b it is
shown a wider portion of phase space and in green the arcs of critical curves
on the boundary of the invariant area, while in Fig.\ref{Unbounded3}c it is
represented the unbounded attracting area in the whole plane, related to the
scaled variables.}

\subsection{Bifurcations occurring in the second enlargement.}

In the previous subsection we have described a few bifurcations which occur
decreasing $h$ and evidenced in Fig.\ref{1DB}b, leading to an unbounded
invariant absorbing area, with complex dynamics. Clearly, inside this
invariant area it is possible that stable cycles of high period exist, not
detectable in our numerical simulations. However, cycles with homoclinic
orbits also exist,{ both saddles and expanding cycles, as the 4-cycle
described above which is a snap-back repellor (in fact, although an invariant
4-cyclic attracting chaotic area no longer exists, as evidenced by the complex
dynamics, the homoclinic 4-cycle still exists).} From Fig.\ref{1DA} we can see
that the unbounded invariant absorbing area persists for a wide interval of
values for the parameter $h$, and the appearance of an attracting 4-cycle
leads to another window with interesting dynamics, enlarged in Fig.\ref{1DB}a.
The pair of 4-cycles appears by saddle-node bifurcation inside the invariant
absorbing area.

\begin{figure}
[h]
\begin{center}
\includegraphics[scale=0.37]{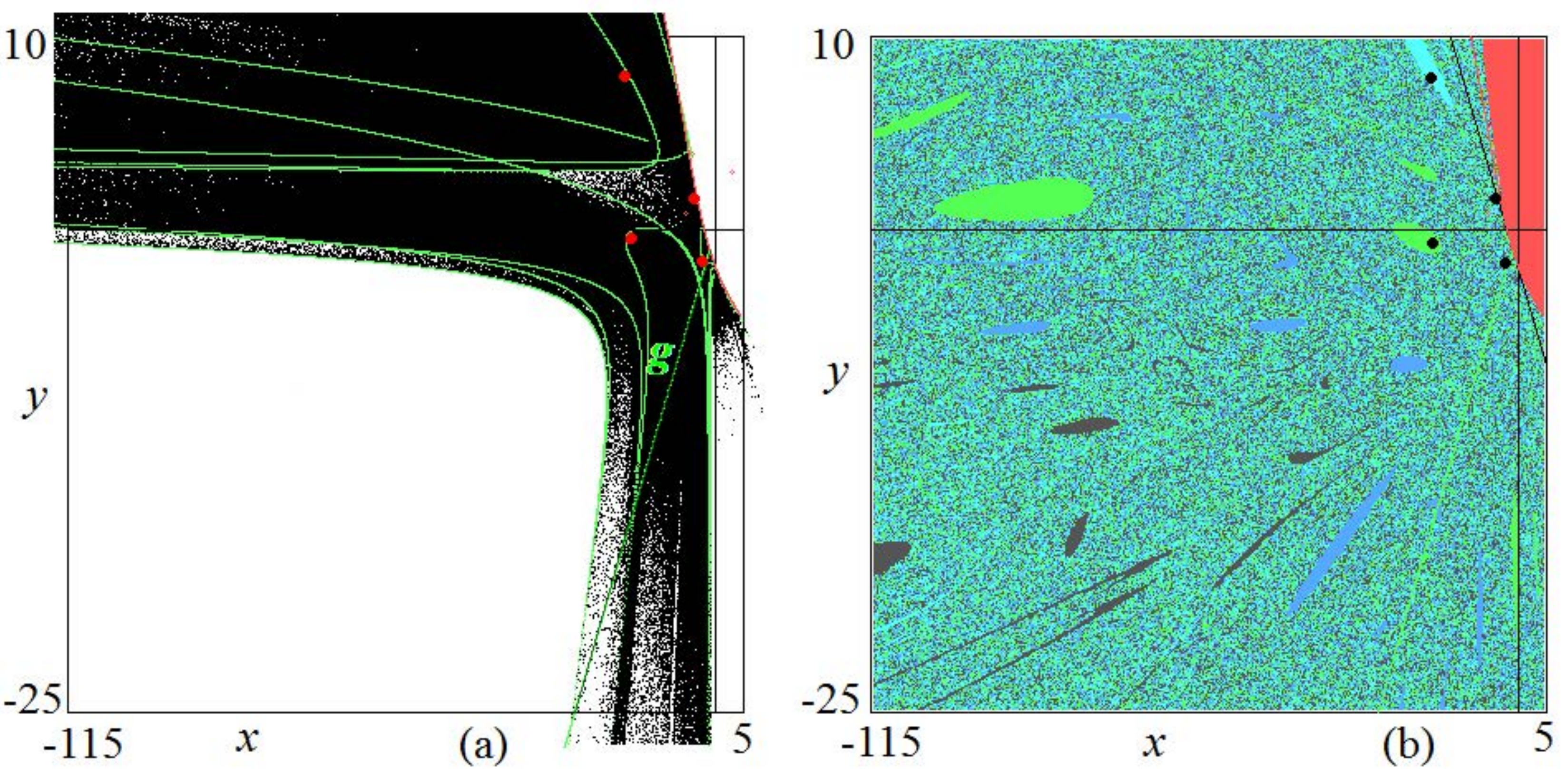}
\caption{Appearance of an attracting 4-cycle. In (a) at $h=5.128$ the cycles
do not exsist. In (b) at $h=5.127$ a pair of 4-cycles exist inside the
unbounded invariant area. }
\label{NewWindow}
\end{center}
\end{figure}

\begin{figure}
[h]
\begin{center}
\includegraphics[scale=0.35]{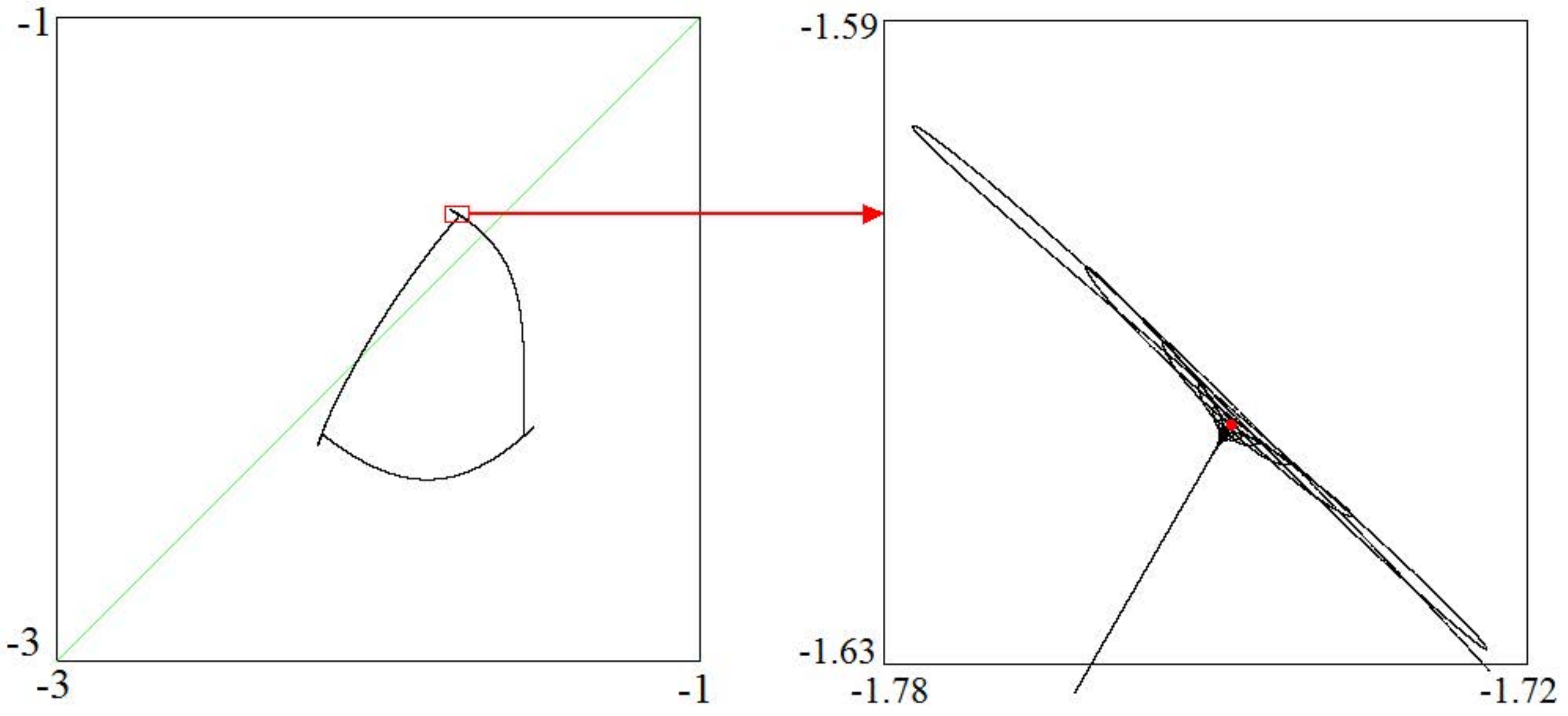}
\caption{Enlargement showing one piece of the 4-cyclic invariant set at
$h=5.10299$ and appearance of the 12-cycle at $h=5.10298$ represented by the
red point in the further enlargement.}
\label{12-cycle}
\end{center}
\end{figure}

In Fig.\ref{NewWindow}a (at $h=5.128$) we show the unbounded invariant area
confined by portions of critical curves, and the generating segment is now a
wider connected segment of $LC_{-1}^{j},$ in red are indicated the points
where a pair of 4-cycles is going to appear. In Fig.\ref{NewWindow}b (at
$h=5.127$) an attracting 4-cycle exists inside the invariant area, and its
basin of attraction is evidenced with respect to the 4 points (in black) which
are fixed for the 4-th iterate of the map $G_{h}$. The complex structure of
the basins evidences a chaotic repellor on the basin boundary, while the
immediate basins are (as in previous examples) bounded by the stable set of
the saddle 4-cycle forming closed curves, denoting intersection with the
critical curves.
From Fig.\ref{1DB}a we can see that the 4-cycle undergoes a $NS$ bifurcation
(at $h\simeq5.11$), leading to 4-cyclic closed invariant curves which, from
smooth (for $h$ close to the $NS$ bifurcation)\ become nonsmooth after the
crossing of the {critical line }$LC_{-1}^{j}$. We also see in Fig.\ref{1DB}a
that an attracting 12-cycle appears, which is related to the closed curves. In
fact, at $h\simeq5.10299$ the attracting set is no longer a smooth curve,
since it is an invariant set with self intersections, as reported in
Fig.\ref{12-cycle} (for {the mechanism leading to the formation of loops and
self intersections in closed curves }see { \cite{FrouzakisEtAl} and \cite{MGBC
96}).}

\begin{figure}
[h]
\begin{center}
\includegraphics[scale=0.32]{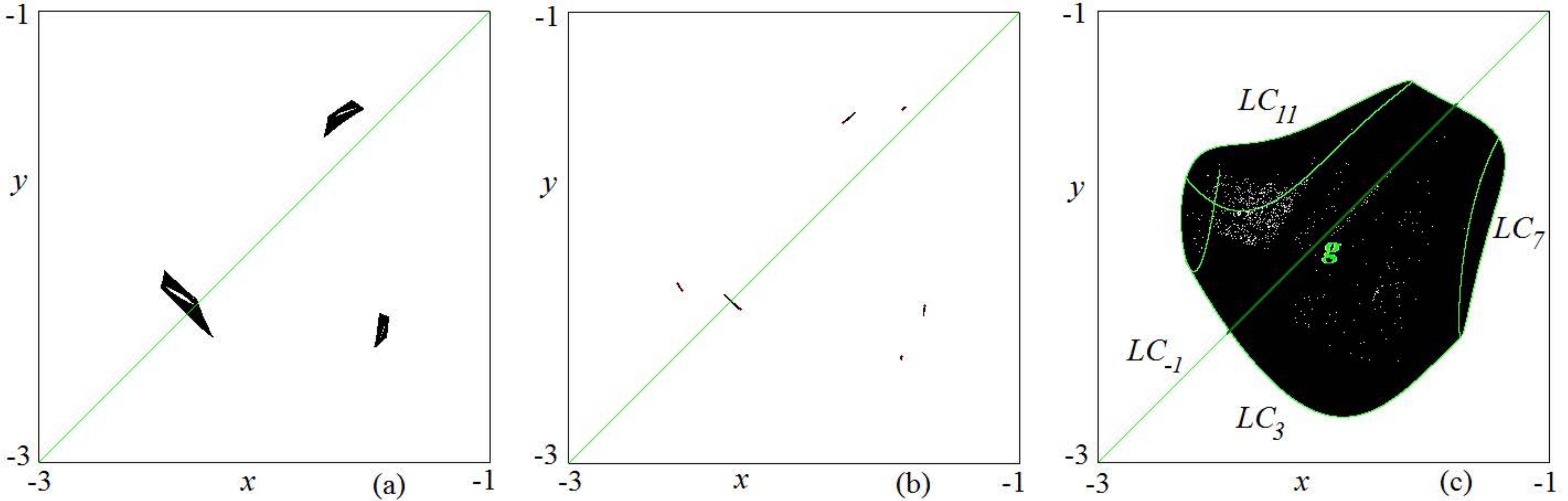}
\caption{Enlargements showing a part of a chaotic attractor. Three pieces of
an attractor of 12 connected components for $h=5.097$ in (a); six pieces
belonging to an invariant set of 24 connected components for $h=5.096$ in (b);
and one of 4 connected components for $h=5.095$ in (c).}
\label{chaos-4}
\end{center}
\end{figure}

From Fig.\ref{1DB}a we also see that, in its turn, the attracting 12-cycle
undergoes a $NS$ bifurcation, and after a sequence of bifurcations 12-cyclic
chaotic pieces are created (see Fig.\ref{chaos-4}a at $h=5.097).$ Other
bifurcations are illustrated in Fig.\ref{chaos-4}b at $h=5.096$ (enlarged
portion of phase plane, evidencing one piece of the {12-cyclic set}). {This,
in its turn, is followed by a contact bifurcation at $h=5.095,$ leading to an
expansion of the invariant absorbing area with chaotic dynamics which has now
$4$-cyclic chaotic pieces, one of them shown in Fig.\ref{chaos-4}c. This also
denotes} that the repelling 4-cycle inside that area is now homoclinic, i.e. a
snap-back repellor.

\begin{figure}
[h]
\begin{center}
\includegraphics[scale=0.37]
{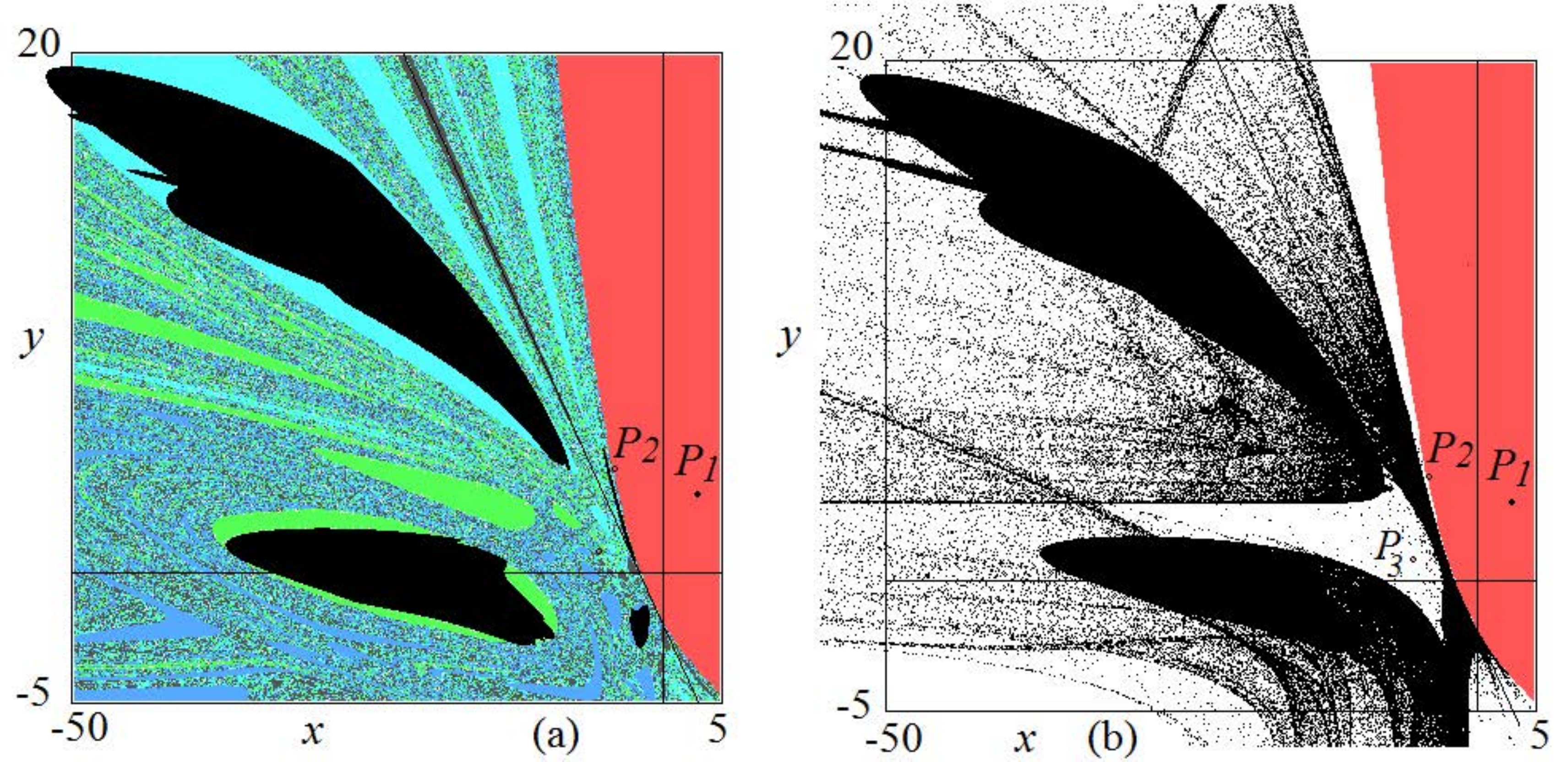}
\caption{Contact bifurcation leading to expansion of the invariant area. In
(a) $h=5.09397$ 4-cyclic absorbing areas exist. In (b) $h=5.09396$ the
attracting set is an unbounded absorbing area.\ }
\label{4-final}
\end{center}
\end{figure}

It is worth noting that all these attracting sets belong to the unbounded
invariant absorbing area, generated by the segment as shown in
Fig.\ref{NewWindow}a. However, inside this region, we can find other invariant
absorbing areas as the 4-cyclic areas which are all bounded by few images of
the generating segment intersecting $LC_{-1}^{j}$ and shown in
Fig.\ref{chaos-4}c.
The 4-cyclic chaotic areas are shown at a lower value of $h$ in
Fig.\ref{4-final}a, together with the related four basins of attraction for
the 4-th iterate of the map. Besides a fractal structure with a chaotic
repellor on the boundary, this figure shows that the areas are very close to
the boundary of their immediate basins, denoting that the parameter is very
close to a contact bifurcation value, leading to the disappearance of the
4-cyclic areas, and the appearance of a unique unbounded absorbing area inside
which the dynamics\ seem chaotic, see Fig.\ref{4-final}b.

The unbounded absorbing area previously obtained, persists for lower values of
$h$, as shown in Fig.\ref{1DA}, up to the bifurcation described in
Sec.\ref{SS241} occurring at $h=\widetilde{h}.$ Decreasing $h$ up to the final
bifurcation, it seems that the dynamics inside the unbounded absorbing area
become more complex. That is, more cycles appear and undergo homoclinic
bifurcations. For example the fixed point $P_{3}$ which is a repelling focus
may become a snap-back repellor. From Fig.\ref{4-final}b it is not clear
whether it is homoclinic or not. However, we have evidence of this, at the
same value $h=5.09396$ used in Fig.\ref{4-final}b.
\begin{figure}
[h]
\begin{center}
\includegraphics[scale=0.35]
{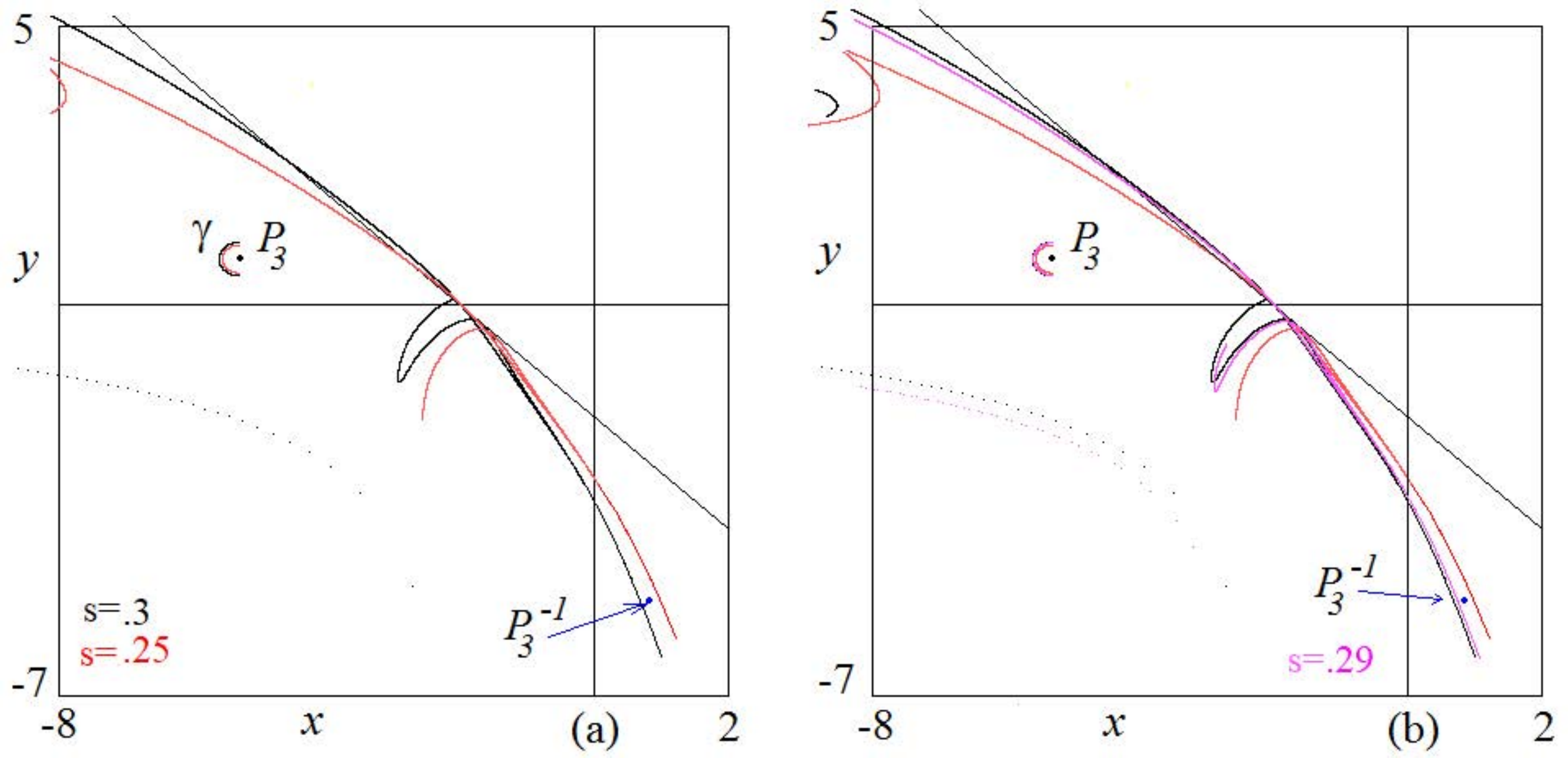}
\caption{Images $G_{h^{\ast}}^{21}(\gamma)$ at $h^{\ast}=5.09396$ of different
arcs $\gamma$ close to $P_{3}$ corresponding to different values of the radius
$s$. In (a) two different arcs are shown. In (b) a third arc is added to the
two shown in (a). The values of $s$ are given in the text.}
\label{gam}
\end{center}
\end{figure}

In fact, let us consider the fixed point $P_{3}$ and its rank-1 preimage
different from itself, $P_{3}^{-1},$ which is the point symmetric to $P_{3}$
with respect to the main diagonal. In order to show that $P_{3}$ is a
snap-back repellor it is enough to show that there is a point $q_{0}$ in a
neighborhood of $P_{3}$ which is mapped into $P_{3}$ in a finite number of
iterations, say $G_{h}^{m}(q_{0})=P_{3},$ and such that the local inverse
which satisfies $G_{h}^{-1}(P_{3})=P_{3}$ (which is a contraction), leads to
$\lim_{n\rightarrow\infty}G_{h}^{-n}(q_{0})=P_{3}.$

Let us consider an arc $\gamma$ in such a neighborhood of $P_{3}$ (i.e. such
that the preimages of its points by the local inverse $G_{h}^{-1}$ converge to
$P_{3}$), given by a half circle of equation
\[
(x-x(P_{3}))^{2}+(y-y(P_{3}))^{2}=s^{2}%
\]
with the radius $s$ sufficiently small. In Fig.\ref{gam}a we show such an arc
$\gamma$ in black corresponding to the radius $s=0.3$, and the image
$G_{h^{\ast}}^{21}(\gamma)$ (with $h^{\ast}=5.09396$ as used in
Fig.\ref{4-final}b) in black which is close to the preimage $P_{3}^{-1},$
below it. Differently an arc $\gamma$ in red corresponding to the radius
$s=0.25$, has the image $G_{h^{\ast}}^{21}(\gamma)$ in red which is close to
the preimage $P_{3}^{-1},$ but above it. This means that there exists a value
$s^{\ast}\in(0.25,0.3)$ such that the related arc $\gamma^{\ast}$ has the
image $G_{h^{\ast}}^{21}(\gamma^{\ast})$ which crosses through $P_{3}^{-1}$.
This implies the existence of a {point $q_{0}\in\gamma^{\ast}$} such that
$G_{h^{\ast}}^{22}(q_{0})=P_{3}.$

As an example, considering $s=0.29$ in Fig.\ref{gam}b we show in pink the
related arc $\gamma$ and, also in pink, the image $G_{h^{\ast}}^{21}(\gamma)$
which is above the black arc and closer to $P_{3}^{-1}.$

A similar construction can be used also at any smaller value of $h$ (in
particular also at $h=5$ or smaller), showing that $P_{3}$ persists in being a snap-back repellor.

\section{Conclusions}

We have considered a one-parameter family of maps which unfolds a Landen-type
map obtained by Boros and Moll, and we have described the main bifurcations
that appear in the route towards the parameter value that corresponds to this
map. We have focused on those homoclinic and contact bifurcations that can
give information about the unbounded invariant chaotic region that appear in
the Boros-Moll map. Our study confirms that the dynamics of this map is
singular in the considered family, and we give evidences that this singularity
is related with a specific property of the critical lines: the fact that one
of the critical lines merges with one separatrix of the saddle point only for
the parameter value corresponding with the map introduced by Boros and Moll.

\bigskip

\subsection*{Acknowledgements}

V.M. is supported by Ministry of Economy, Industry and Competitiveness--State
Research Agency of the Spanish Government through grant DPI2016-77407-P
(MINECO/AEI/FEDER, UE) and acknowledges the group research recognition
2017-SGR-388 from AGAUR, Generalitat de Catalunya. He also acknowledges Prof.
A.~Gasull for helpful discussions. V.M. and I.S. as visiting professor are
grateful to the University of Urbino Carlo Bo for the hospitality during a
stage at the University in February 2018.


\section{Appendix}

The map $G_{h}$ has a non unique inverse, and the number of distinct rank-1
preimages of a point $(x,y)$ differs depending on the region of the phase
plane which the point $(x,y)$ belongs to. These regions, or zones $Z_{i}$
(related to $i$ distinct preimages of rank-1) following the notation
introduced in \cite{MGBC 96}, are usually bounded by critical curves $LC$\ of
the map, which are the images of curves of merging preimages, denoted
$LC_{-1}$, which belong to the set where the Jacobian determinant of the map vanishes.

In our case, the Jacobian determinant vanishes either at the points
$LC_{-1}^{j}$ (defined by $y=x$) or the points in $LC_{-1}^{jj}$ (defined by
$x+y-6=0$), see Eqs. \eqref{E-jacobi} and \eqref{E-LC-1}.

Since $LC_{-1}^{j}$ intersects $(d)$, its image $LC^{j}$ consists of two
unbounded curves of the plane, $LC^{j}\cup LC_{r}^{j}$ (notice that if an arc
$\gamma$ crosses the line $(d)$, where the denominator vanishes in a point in
which the numerators of the components of the map do not vanish, which is
always the case for the considered map, then its image $G_{h}(\gamma)$
consists of two unbounded arcs).
\begin{figure}
[h]
\begin{center}
\includegraphics[scale=0.35]
{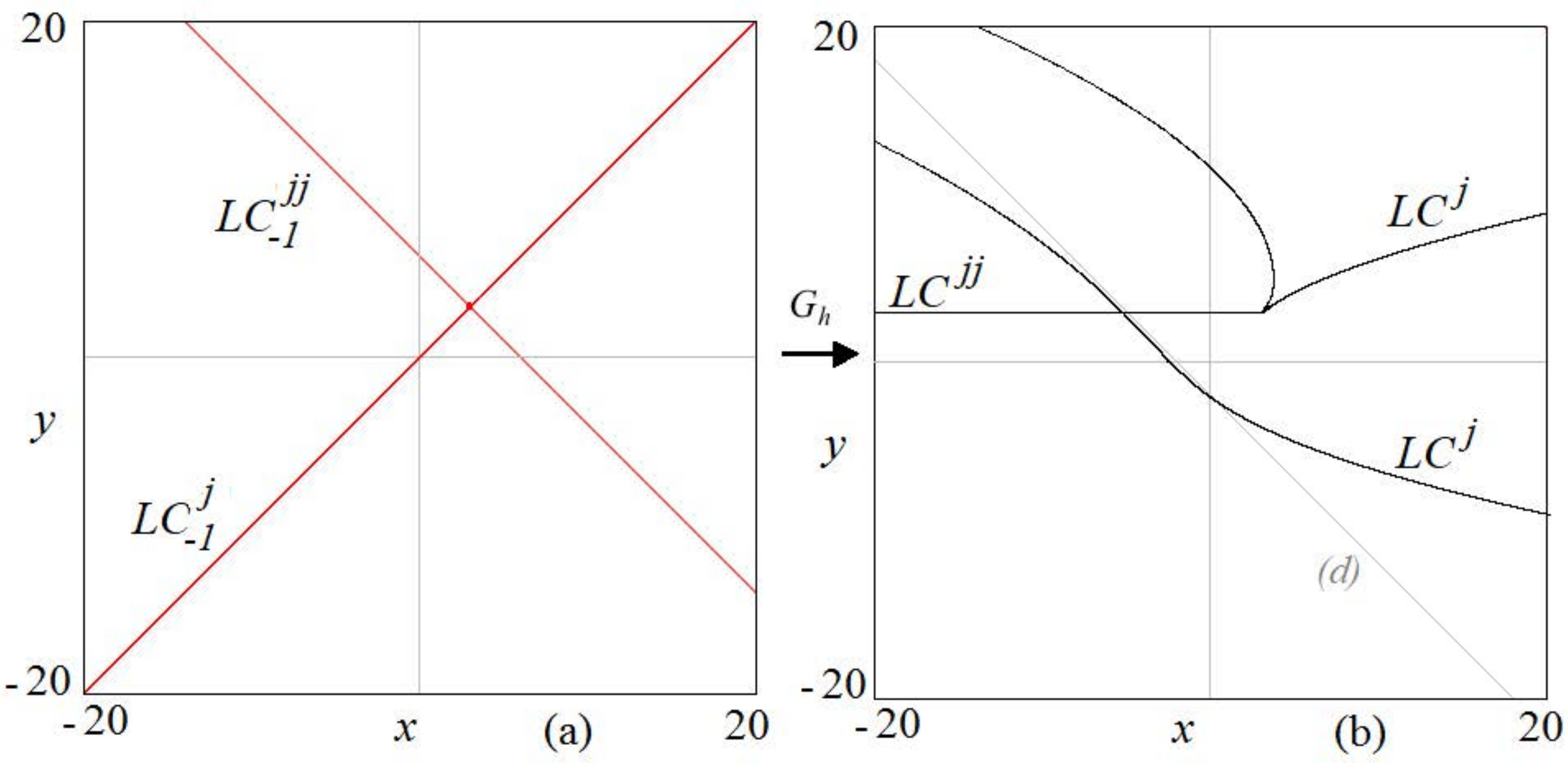}
\caption{Critical curves for the map $G_{h}$ at $h=5.5.$ In (a) $LC_{-1};$ in
(b) $LC$.}
\label{F-cc}
\end{center}
\end{figure}

The curve $LC_{-1}^{jj}$ does not intersects $(d)$, its image $LC^{jj}%
=G_{h}(LC_{-1}^{jj})$ consists of a single arc, and it is easy to see that it
is a segment of straight line $y=3$ (up to $x=3$). Indeed, by setting
$x_{n}=t$ and $y_{n}=-t+6$ (a point on $LC_{-1}^{jj}$), and taking
$(x_{n+1},y_{n+1})=G_{h}(x_{n},y_{n})$ we get $x_{n+1}=(6t-t^{2}+39)/16$
(which takes its maximum value in $x=3$) and $y_{n+1}=3$, that is, the image
of the line\ $LC_{-1}^{jj}$ is folded on the half-line $y=3$, as $t$ increases
from $-\infty$ to $+\infty$ and $x=t=3$\ is the folding point. In
Fig.\ref{F-cc} the critical curves for $h=5.5$ are displayed.

\begin{figure}[h]
\begin{center}
\includegraphics[scale=0.28]
{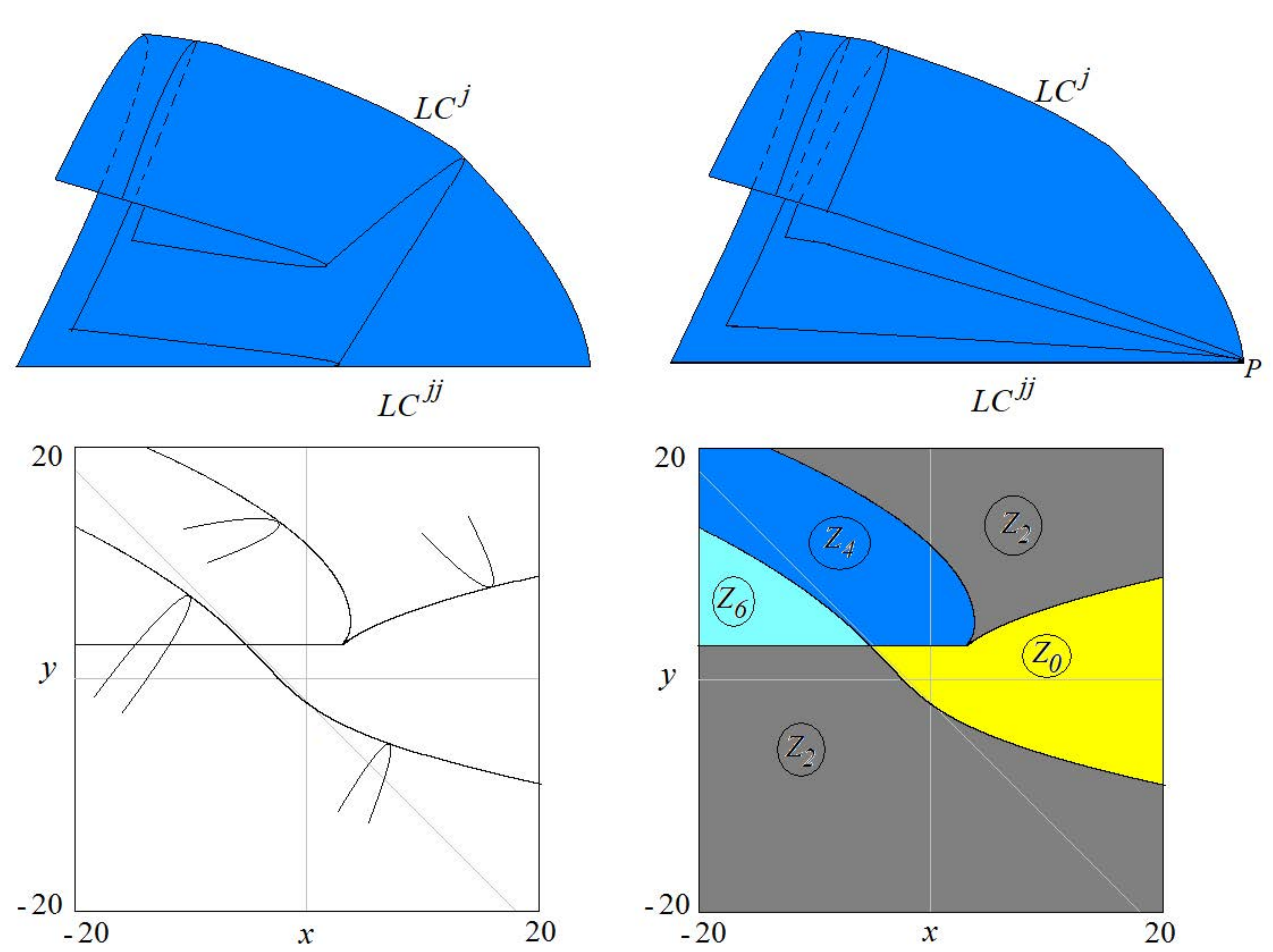}
\caption{Qualitative shape of the foliation of the plane. Critical curves and
Zones $Z_{i}$ for the map $G_{h}$ at $h=5.5.$}
\label{F-foliation}
\end{center}
\end{figure}

In summary, the components of the critical curve $LC=LC^{j}\cup LC_{r}^{j}\cup
LC^{jj}$ split the plane in $5$ distinct open regions. We can figure out the
number of preimages by considering a sort of \textquotedblleft folding of the
plane", as qualitatively shown in Fig.\ref{F-foliation}. By computing how many
preimages have \emph{a particular point} in each of these regions, we obtain
the different zones $Z_{i}$. In Fig.\ref{F-foliation} they are displayed for
the particular case $h=5.5$. Notice also that a point belonging to $LC$ has
two merging rank-1 preimages in a point belonging to $LC_{-1}.$ For each point
in a zone $Z_{i}$ it is also possible to show where the preimages are located
(on opposite side with respect to the related branch of $LC_{-1})$.

We report how to obtain the number of preimages in each of the zones limited
by the curves $LC$. We fix a point $(u,v)$ and we set $G_{h}(x,y)=(u,v)$. By
introducing the notation $z=x+y+2$ and from the equation of the second
component of the map, we have that $vz^{2/3}=z+4$, hence
\begin{equation}
v^{3}={(z+4)^{3}}/{z^{2}}\ . \label{E-v3z}%
\end{equation}
By studying the function $f(z)={(z+4)^{3}}/{z^{2}}$ we easily get that for
$v>3$ there are three solutions of Eq. \eqref{E-v3z} (one negative and two
positive); for $v=3$ there is one negative solution and two merging solutions
given by $z=8$; and for $-\infty<v<3$ there is only one negative solution.

Introducing the notation $P=xy$ and $S=x+y$, and by using the equation of the
first component, we obtain that given $(u,v)$ and a value of $z$ satisfying
Eq. \eqref{E-v3z}, there is a unique value for $P$ and $S$ given by
\[
P=u(z^{4})^{1/3}-h(z-2)-9\,\mbox{ and }S=z-2\ .
\]
Since any preimage $(x,y)$ associated to $(u,v)$ satisfies $y^{2}-Sy+P=0$ and
$x=S-y$, we have
\[
x=\frac{S\pm\sqrt{S^{2}-4P}}{2}\,\mbox{ and }y=\frac{S\mp\sqrt{S^{2}-4P}}%
{2}\ .
\]
The final number of preimages depends on the sign of $S^{2}-4P$.

\begin{figure}[h]
\begin{center}
\includegraphics[scale=0.27]
{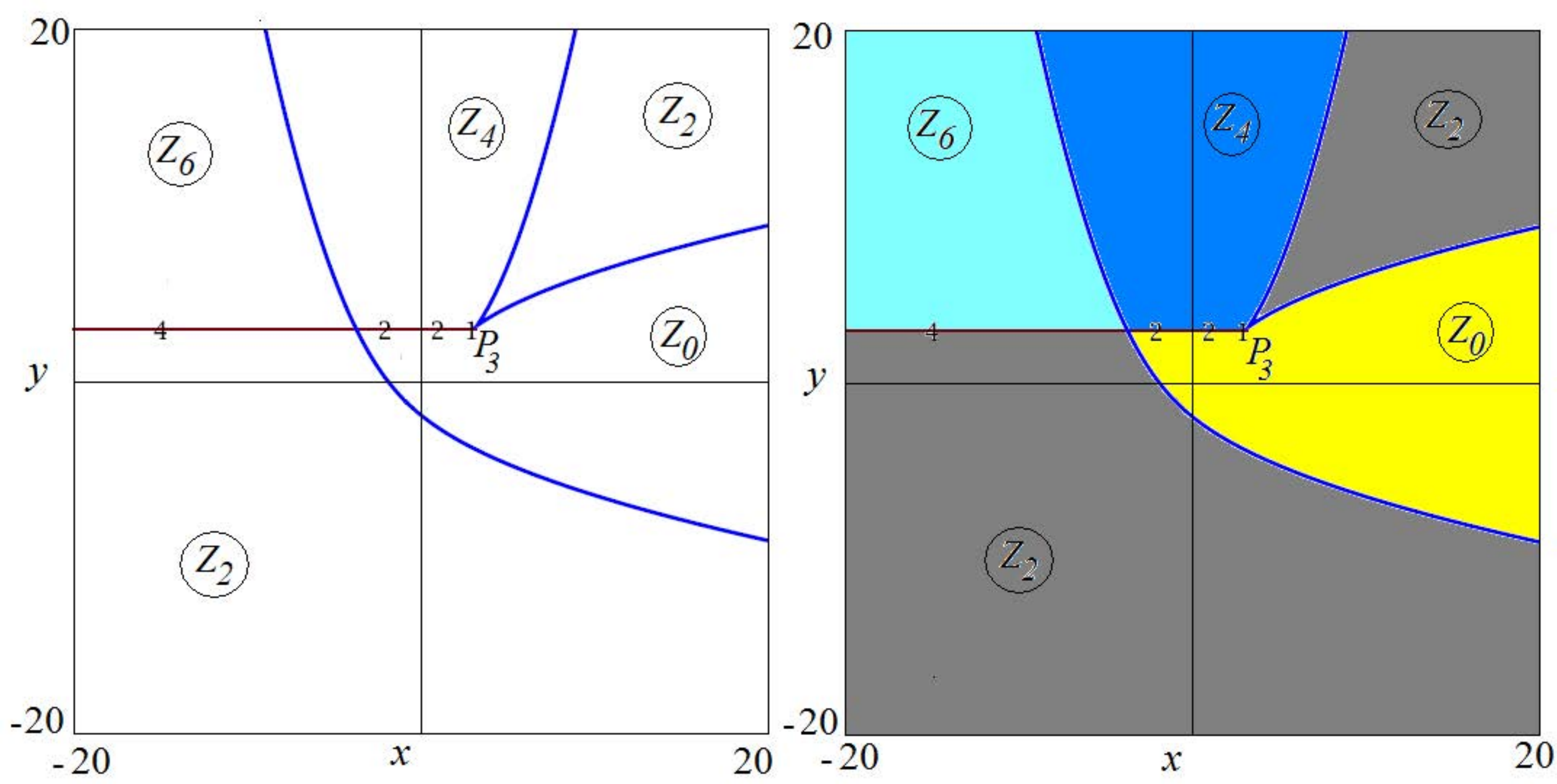}
\caption{Critical curves and Zones for the Boros-Moll map ($h=5$).}
\label{F-preimMoll}
\end{center}
\end{figure}

For the original Boros-Moll map, the case $h=5$, the curve $LC^{j}$ belongs to
the curve $R(x,y)=-{x}^{2}{y}^{2}+4\,{x}^{3}+4\,{y}^{3}-18\,xy+27=0$, it is
part of the stable set of the saddle $P_{2}$, and the zones $Z_{i}$ are
displayed in Figure \ref{F-preimMoll}.

\end{document}